\documentclass[reprint,amsmath,amssymb,epjb]{revtex4-2}
\usepackage[caption=false]{subfig}
\usepackage{listings}
\usepackage{xcolor}
\usepackage{multirow}
\usepackage{graphicx}
\usepackage{dcolumn}% Align table columns on decimal point
\usepackage{bm}
\usepackage{braket}
\makeatletter
\newcommand{\vast}{\bBigg@{3}}
\newcommand{\Vast}{\bBigg@{4}}
\begin{document}
\title{Nash equilibrium mapping vs. Hamiltonian dynamics vs. Darwinian evolution for some social dilemma games in the thermodynamic limit}
\author{Colin Benjamin}
\email{colin.nano@gmail.com}
\author{Arjun Krishnan U M}
\affiliation{School of Physical Sciences, National Institute of Science Education and Research, Jatni 752050, India}
\affiliation{Homi Bhabha National Institute, Training School Complex, Anushaktinagar, Mumbai 400094, India}
\begin{abstract}
How cooperation evolves and manifests itself in the thermodynamic or infinite player limit of social dilemma games is a matter of intense speculation. Various analytical methods have been proposed to analyze the thermodynamic limit of social dilemmas. In this work, we compare two analytical methods, i.e., Darwinian evolution and Nash equilibrium mapping, with a numerical agent-based approach. For completeness, we also give results for another analytical method, Hamiltonian dynamics. In contrast to Hamiltonian dynamics, which involves the maximization of payoffs of all individuals, in Darwinian evolution, the payoff of a single player is maximized with respect to its interaction with the nearest neighbour. While the Hamiltonian dynamics method utterly fails as compared to Nash equilibrium mapping, the Darwinian evolution method gives a false positive for game magnetization- the net difference between the fraction of cooperators and defectors- when payoffs obey the condition $a+d=b+c$, wherein $a$,$d$ represent the diagonal elements and $b$,$c$ the off-diagonal elements in a symmetric social dilemma game payoff matrix. When either $a+d \neq b+c$ or when one looks at the average payoff per player, the Darwinian evolution method fails, much like the Hamiltonian dynamics approach. On the other hand, the Nash equilibrium mapping and numerical agent-based method agree well for both game magnetization and average payoff per player for the social dilemmas in question, i.e., the Hawk-Dove game and the Public goods game. This paper thus brings to light the inconsistency of the Darwinian evolution method vis-a-vis both Nash equilibrium mapping and a numerical agent-based approach.
\end{abstract}
\maketitle

\section{Introduction}
This paper aims to compare the three analytical methods developed using equilibrium statistical mechanics in analogy with the $1D$ Ising model: Nash equilibrium mapping, Darwinian evolution, and Hamiltonian dynamics. Our idea is to check which model is closest in predicting the Nash equilibrium behaviour of players in the thermodynamic limit while subject to a temperature or noise (a measure of randomness in the choice of strategy) denoted by $\beta=1/k_BT$. Finding the ideal method is useful for studying how an infinite number of players interact in social dilemmas. We try to understand cooperative behaviour in the thermodynamic limit using the game magnetization and average payoff per player as indicators. The numerical agent-based approach agrees with Nash Equilibrium mapping method, regardless of whether payoffs obey condition $a+d=b+c$ or not, where $a,b,c,d$ represent the payoffs associated with strategy tuples $(s_1,s_1),(s_1,s_2),(s_2,s_1),(s_2,s_2)$ respectively in a symmetric 2-player 2-strategy game matrix ($s_1$ and $s_2$) represent the two strategies available to each player). However, when $a+d \neq b+c$ results obtained for game magnetization and average payoff per player via the numerical agent-based approach completely disagree with the Darwinian Evolution model. Hamiltonian Dynamics model, by definition, cannot be applied to games that do not obey payoff condition $a+d = b+c$. In this paper, we first deal with the Hawk-Dove game where payoff condition $a+d = b+c$ is never satisfied, and then we discuss the Public goods game wherein this condition is always satisfied. So, unlike a game such as the Prisoner's Dilemma, which can accommodate both $a+d = b+c$ as well as $a+d \neq b+c$ payoff conditions in its payoff matrix, we study two games that exclusively allow either of these payoff conditions, $a+d=b+c$ (Public goods) or $a+d\neq b+c$ (Hawk-Dove). 

An analogy to the $1D$ Ising model has been invoked to understand the Nash equilibrium strategy of a social dilemma game in the limit where there are infinite players in the system, i.e., thermodynamic limit~\cite{Colin, Colin2, Adami}. Spin sites correspond to players, and their spin states correspond to cooperate or defect strategies~\cite{Colin}. Social dilemmas are addressed in the thermodynamic limit for 2-player, two-strategy games wherein each player interacts with its nearest neighbour, akin to spin site interaction in the $1D$ Ising model. One such approach is known as the Hamiltonian dynamics (HD) model. The analogy with a $1D$ Ising spin chain has been used to derive game magnetization, which is the net difference between the fraction of cooperators and defectors in the system, similar to how magnetization in the Ising model is defined as the difference between the fraction of up and down spins~\cite{Adami}. However, the HD model gave incorrect game magnetization and average payoff per player in the zero noise or, $\beta\rightarrow \infty$ limit~\cite{Colin}. Also, the game magnetization was not in agreement with the Nash equilibrium strategy of the game, as pointed out by one among us in Refs.~\cite{Colin, Colin2}. It is because the HD model~\cite{Adami,Raleg} inherently attempts to minimize the energy of the whole system, i.e., tries to maximize the collective payoff of all players.

On the other hand, Refs.~\cite{Colin, Colin2} introduces Nash equilibrium (NE) mapping, which maps payoffs in the social dilemma game to the Hamiltonian of the two-site Ising model wherein spin sites interact with only their nearest neighbours. Magnetization of social dilemma game calculated using NE mapping approach, in the thermodynamic limit, is then defined in terms of magnetization of $1D$ Ising model with coupling constant and magnetic field derived in terms of game payoffs. It should be noted that the results for the Nash equilibrium mapping model are consistent with the Nash equilibrium strategy of the social dilemma game in question.

In this paper, for the first time we compare the two analytical methods: NE mapping and Darwinian evolution (DE)~\cite{Adami}. For completeness, we also include results from the HD model which was dealt with earlier in Refs.~\cite{Colin,Colin2}. In the DE model, a focal player is chosen, and its payoff is maximized with respect to its neighbours. The payoff of the focal player is chosen to model energies of spin sites in the Ising chain, and in the DE model, these are written as negative payoffs of the social dilemma game. The energy of a single spin site is minimized with respect to its nearest neighbour. The average payoff per player in zero noise, or $\beta\rightarrow \infty$ limit for Nash equilibrium mapping, agrees with Nash equilibrium payoff of the social dilemma game when payoffs obey the condition $a+d=b+c$, where $a,b,c$ and $d$ represent the payoffs in the symmetric 2-player, 2-strategy game matrix: $U=\left(\begin{array}{c c} a, a & b,c \\ c,b & d,d \end{array}\right)$. Game magnetization for the DE model and NE mapping compare favorably when game payoffs obey the condition $a+d=b+c$. However, the average payoff in the zero noise or $\beta\rightarrow \infty$ limit for the DE model disagrees with the payoff corresponding to the Nash equilibrium strategy regardless of whether condition $a+d=b+c$ is obeyed in the Hawk-Dove game where payoff condition $a+d=b+c$ can never hold, NE mapping and DE model disagree for both game magnetization and average payoff per player in zero noise limit. Besides the three analytical methods, we include a numerical agent-based approach~\cite{Adami}.

The following sections will review the theoretical framework concerning NE mapping, HD and DE models, and an agent-based simulation. Next, we will study these models when applied first to the Hawk-Dove game and then to the Public goods game. We will also analyze and summarize our findings and study how these models match or differ in the case of each game. Finally, we conclude with two tables discussing the Hawk-Dove game and Public goods game wherein NE mapping agrees with agent-based simulation for all cases, but HD and DE models do not. The paper ends with an appendix detailing the code we used for Agent-based simulation.
\section{Theory}
This section explains the three analytical methods that use the $1D$ Ising model analogy to study social dilemma games in the thermodynamic limit. These are namely Nash equilibrium (NE) mapping, Hamiltonian dynamics (HD), and Darwinian evolution (DE) model. The analytical work is based on equilibrium statistical mechanics. Further, we describe a numerical agent-based method (ABM) to simulate games with many players, i.e., thermodynamic limit.

Our work is not focused on a dynamic process of strategy selection, but rather, we look at the equilibrium behaviour of players in the thermodynamic limit. When looking at a spatial model where players are linked to each other via Ising-type coupling, the temperature measures uncertainty/noise where a player makes a random choice between strategies. It implies randomness in the choice of strategy. In nature, an example of this is the process of genetic drift, where random genes survive over those with higher survivability (or, higher fitness) due to pure chance. $\beta$ has also been complimentarily interpreted as the strength of selection (see Ref.~\cite{altrock}), where transition probability for switching strategies (given that the payoff increases) asymptotically goes from zero to $1$ as $\beta$ goes from zero to infinity.

Initially, literature studying evolutionary game theory was approached using non-equilibrium dynamics. The equilibrium statistical mechanics approach was pioneered by Ref.~\cite{Adami}, where the authors introduced HD and DE models. One among us, intrigued by the novelty of this approach, introduced NE mapping, which also uses equilibrium statistical mechanics as its crutch, and showed how it gives better results than the HD model in Refs.~\cite{Colin, Colin2}. In this paper, we compare these models by studying Hawk-Dove and public goods games and prove the incorrectness of the DE model vis-à-vis NE mapping. The motivation of this manuscript is not to look at the time-dependent dynamics of the evolution of strategies but rather to invoke the mathematics of equilibrium statistical mechanics by drawing comparisons with the solution of the $1D$ Ising model. We do this to see how the results would look and to compare them with the Nash equilibrium of the games in question. Our model does not present a system that evolves its strategy over time. Rather, it presents a picture of a one-shot game, where an infinite number of players decide their strategies in one go while subject to noise (temperature) denoted by $\beta$.

\subsection{\label{Isingmodeltheory}1D Ising model}
The Hamiltonian for $1D$ Ising model with N sites where each site interacts with only its nearest neighbour via a symmetric exchange coupling $J$ and global external magnetic field $h$ is given by,
\begin{equation}
H=-J\sum_{i=1}^N\sigma_i\sigma_{i+1}-h\sum_{i=1}^N\sigma_i,
\label{eqn:NEHam}
\end{equation}
where $\sigma=+1,-1$ denotes up and down spins, respectively. The partition function for this system~(\ref{eqn:NEHam}) can be written as,
\begin{equation}
Z=\sum_{\sigma_1}\sum_{\sigma_2}...\sum_{\sigma_N}e^{\beta(J\sum_{i=1}^N\sigma_i\sigma_{i+1}+\frac{h}{2}\sum_{i=1}^N(\sigma_i+\sigma_{i+1})}.
\label{Z}
\end{equation}
Here, $\beta=1/(k_BT)$. To solve~(\ref{Z}), we define a transfer matrix $T$ as follows~\cite{Ising},
\begin{eqnarray}
T(\sigma,\sigma')&=&\bra{\sigma}T\ket{\sigma'}=e^{\beta(J\sigma \sigma' + \frac{h}{2}(\sigma + \sigma'))},\hspace{5mm} \nonumber\\
\mbox{or, }T&=&\begin{pmatrix}
e^{\beta(J+h)} & e^{-\beta J} \\
e^{-\beta J} & e^{\beta(J-h)}
\end{pmatrix}.
\end{eqnarray}
The partition function can then be written as,
\begin{align}
Z=&\sum_{\sigma_1}\sum_{\sigma_2}...\sum_{\sigma_N}T(\sigma_1,\sigma_2)T(\sigma_2,\sigma_3)...T(\sigma_N,\sigma_1) \notag \\
=&Tr(T^N).
\end{align}
Trace of a matrix is given by sum of eigenvalues. Let's say $\lambda_+,\lambda_-$ are eigenvalues of $T$. This implies $\lambda_+^N,\lambda_-^N$ are respectively eigenvalues of $T^N$. Thus, $Z=Tr(T^N)=\lambda_+^N +\lambda_-^N$. Eigenvalues for $T$ matrix are,
\begin{equation}
\lambda_{+,-}=e^{\beta J}[\cosh(\beta h)\pm \sqrt{\sinh^2(\beta h)+e^{-4\beta J}}].
\end{equation}
Thus, partition function
\begin{equation}
Z=Tr(T^N)=\lambda_+^N +\lambda_-^N=\lambda_+^N[1+(\frac{\lambda_-}{\lambda_+})^N],
\end{equation}
and as $\lambda_-<\lambda_+$, in the thermodynamic limit, we have $(\frac{\lambda_-}{\lambda_+})^N \xrightarrow{} 0$ for $N \xrightarrow{} \infty $, and partition function,
\begin{equation}
Z=\lambda_+^N = e^{N\beta J}[\cosh(\beta h)+\sqrt{\sinh^2(\beta h)+e^{-4\beta J}}]^N.
\end{equation}
Free energy of the system $F=-k_BT\ln(Z)$. Thus, magnetization per site, which is the difference between the fraction of spins pointing up and the fraction of spins pointing down, can be obtained from Free energy as
\begin{equation}
m=-\frac{1}{N}\frac{dF}{dh}=\frac{1}{N}\frac{1}{\beta}\frac{1}{Z}\frac{dZ}{dh}=\frac{\sinh(\beta h)}{\sqrt{\sinh^2(\beta h)+e^{-4\beta J}}}.
\end{equation}
\subsection{\label{NEtheory}Nash equilibrium (NE) mapping}
To understand the NE mapping approach in the thermodynamic limit of social dilemmas, we start with the $1D$ Ising model with two spin sites. Hamiltonian for such a $2$-site system is,
\begin{equation}
H=-J(\sigma_1\sigma_2+\sigma_2\sigma_1)-h(\sigma_1+\sigma_2).
\end{equation}
The energy for each spin site can be written as
\begin{equation}
E_1=-J\sigma_1\sigma_2-h\sigma_1\hspace{5mm}\text{ and }\hspace{5mm}E_2=-J\sigma_2\sigma_1-h\sigma_2.
\end{equation}
In game theory, players search for Nash equilibrium with maximum possible payoffs; in the Ising model, the equilibrium is defined as minimum energy. Therefore, to draw an equivalence between energy equilibrium which is a minimum, and the Nash equilibrium, which is the best possible (or, maximum possible) payoffs, we write the energy matrix with negative energies $-E_i$ with respect to spins $\sigma_i = +1,-1$, $i=1,2$. $\sigma_1, \sigma_2$ are spins of the first and second sites, respectively. Thus,
\begin{equation}
-E= \Vast(\begin{array}{c|c c}
& \sigma_2=+1 & \sigma_2=-1 \\
\hline
\sigma_1=+1 & J+h,J+h & -J+h,-J-h \\
\sigma_1=-1 & -J-h,-J+h & J-h,J-h
\label{eqn:Isingmatrix}
\end{array}\Vast).
\end{equation}
Now a general symmetric payoff matrix for a two-player game is given as,
\begin{equation}
U=
\Vast(\begin{array}{c|c c}
& s_1 & s_2 \\
\hline
s_1 & a,a & b,c \\
s_2 & c,b & d,d
\end{array}\Vast).
\label{eqn:abcdmatrix}
\end{equation}
Introducing a transformation to the general payoff matrix~(\ref{eqn:abcdmatrix}) such that Nash equilibrium remains the same~\cite{Colin3}, and a one-to-one mapping between transformed payoff matrix and energy payoff matrix depicted in Eq.~(\ref{eqn:Isingmatrix}) can be brought about. {We transform the payoff's $a, b, c, d$ to $a+\lambda, b+\mu, c+\lambda, d+\mu$, where $\lambda, \mu$ represent the transformation parameters.} Choosing $\lambda=-\frac{a+c}{2}$ and $\mu=-\frac{b+d}{2}$, we get
\begin{equation}
U'=\Vast(\begin{array}{c|c c}
& s_1 & s_2 \\
\hline
s_1 & \frac{a-c}{2},\frac{a-c}{2} & \frac{b-d}{2},\frac{c-a}{2} \\
s_2 & \frac{c-a}{2},\frac{b-d}{2} & \frac{d-b}{2},\frac{d-b}{2}
\end{array}\Vast).
\label{eqn:transformedpayoffmatrix}
\end{equation}
Note that such a mapping preserves the Nash equilibrium of the game~\cite{galam}. Equating each matrix element of Eq.~(\ref{eqn:transformedpayoffmatrix}) with that of Eq.~(\ref{eqn:Isingmatrix}) provides a relation between payoffs and Ising parameters as,
\begin{equation}
J=\frac{a-c+d-b}{4}\hspace{5mm} \text{ and }\hspace{5mm}h=\frac{a-c+b-d}{4}.
\label{eqn:Jandh}
\end{equation}
Thus, in analogy to magnetization for a spin, we can define a game magnetization $m_g$ (difference between the fraction of people using strategy $s_1$ and those using strategy $s_2$) for games~(\ref{eqn:abcdmatrix}) in thermodynamic limit (for payoffs $a+d\neq b+c$),
\begingroup
\begin{equation}
m_g=\frac{\sinh \beta (\frac{a-c+b-d}{4})}{\sqrt{ \sinh^2\beta( \frac{a-c+b-d}{4})+e^{-4\beta(\frac{a-c+d-b}{4})}}}.
\label{eqn:magNEmappingabcdgeneral}
\end{equation}
\endgroup
In the special case of a game obeying payoff condition $a+d=b+c$, we have $J=0$ and $h=(a-c)/2$. Thus, game magnetization, in this case, becomes,
\begin{equation}
m_g=\tanh \beta \bigg( \frac{a-c}{2}\bigg).
\label{magNEmappingabcdwithcondition}
\end{equation}

The temperature $T$ is a measure of noise or randomness in strategy. Increasing $\beta=1/(k_BT)$ (i.e., decreasing noise) leads to less randomness in the choice of strategy. For $\beta\rightarrow\infty$, noise vanishes, and the system is at Nash equilibrium, while in the $\beta\rightarrow 0$ limit, strategy choices are completely randomized due to maximum noise. Hence, $\beta$ acts to randomize strategic choices. In evolutionary theory, $\beta$ is the equivalent of the rate at which fitness fluctuates due to different strengths of selection (e.g., genetic drift and mutations), i.e., selection intensity\cite{altrock}.
\subsection{\label{HDtheory}Hamiltonian Dynamics (HD) model}
HD model is an analytical framework to study social dilemma games using the analogy with the $1D$ Ising model, similar to NE mapping. The nomenclature is taken from Ref.~\cite{Adami}, where this model was first introduced. It is a misnomer since they use equilibrium statistical mechanics for their mathematics and are not concerned with how a system of players develops over time (as we also do). However, we are sticking with their nomenclature in our work, too, so there would be no confusion for the readers when they go through the reference literature.

However, there are important differences between NE mapping and the HD model. The spin states are equivalent to strategies the players adopt in a game and are represented here as ket vectors. Spin upstate is the equivalent of cooperate strategy and is represented as $\ket{0}=\begin{pmatrix}1\\0\\ \end{pmatrix}$ and spin down is the equivalent of defect strategy and is given by $\ket{1}=\begin{pmatrix}0\\1\\ \end{pmatrix}$. The state of the system with $N$ spin sites is the direct product of individual spin vectors of all the sites. This is written as $\ket{x}=\ket{m_1m_2m_3....m_N},m_i \in \{ 0,1 \}$. Since we are usually concerned with symmetric games, the payoff matrix $U$ represents the row player's payoffs, and the payoffs received by the column player can be deduced from the same.
\begin{equation}
U=
\Vast(\begin{array}{c|c c}
& s_1 & s_2 \\
\hline
s_1 & a & b \\
s_2 & c & d
\end{array}\Vast).
\end{equation}
We introduce the energy matrix $E$ as the negative of the game payoff matrix. It is this energy matrix that we will be using to define the Hamiltonian. The energy values are chosen as negative game payoffs because, in the game, the players always try to maximize their payoff. At the same time, a physical system strives to minimize its energy.
\begin{equation}
E=\begin{pmatrix}
E_{00} & E_{01}\\
E_{10} & E_{11}
\end{pmatrix}
=\begin{pmatrix}
-a & -b\\
-c & -d
\end{pmatrix}
\label{payoffandenergymatrix}
\end{equation}
Hamiltonian for the system where each player interacts with its nearest neighbour is given by (see, Ref.~\cite{Adami}),
\begin{equation}
H=\sum_{i=1}^N\sum_{m,n=0}^1E_{mn}P_m^{(i)}\otimes P_n^{(i+1)}.
\label{eqn:Ham}
\end{equation}
In Eq.~(\ref{eqn:Ham}), $P_0$ and $P_1$ are projectors defined as $P_0=\ket{0}\bra{0}$ and $P_1=\ket{1}\bra{1}$. $P_0^{(i)}$ and $P_1^{(i)}$ denote projection operators for up spin (cooperation) and down spin (defection), respectively, for spin site $i$, while $m,n \in \{0,1\}$ denote indices of energy matrix $E$. Ref.~\cite{Adami} proceeds to find game magnetization for the system using the HD approach. The problems with this approach have been dealt with extensively in Refs.\cite{Colin, Colin2}. Still, in this work, for completeness, we discuss the HD approach, derive both game magnetization and average payoff per player, and use it to compare results obtained with other models. We start by writing the partition function for Hamiltonian depicted in (\ref{eqn:Ham}) as
\begin{align}
Z=&Tr(e^{-\beta H})=\sum_{\ket{x}}\bra{x}e^{-\beta H}\ket{x}, \notag\\
=&\sum_{\ket{m_1m_2...m_N}}\bra{m_1m_2...m_N}e^{-\beta H}\ket{m_1m_2...m_N}.
\end{align}
Taking the expectation value of $H$, we find
\begin{equation}
\bra{m_1...m_N}H\ket{m_1...m_N}=E_{m_1m_2}+E_{m_2m_3}+...+E_{m_Nm_1}.
\end{equation}
For mathematical convenience, we define a matrix $K$ with elements $K_{i j}=e^{-\beta E_{i j}}$,
\begin{equation}
\text{Thus, }K=\begin{pmatrix}
e^{\beta a} & e^{\beta b} \\
e^{\beta c} & e^{\beta d}
\end{pmatrix}.
\label{K}
\end{equation}
The partition function can then be written as,
\begin{align}
Z=&\sum_{m_1,m_2...m_N}e^{-\beta(E_{m_1m_2}+E_{m_2m_3}+...+E_{m_Nm_1})}, \notag\\
=&\sum_{m_1m_2...m_N}K_{m_1m_2}K_{m_2m_3}...K_{m_Nm_1}.
\end{align}
One can see this is the formula that gives a trace of the matrix $K$ raised to its $N^{th}$ power,
\begin{equation}
Z=Tr(K^N).
\label{partitionfunctionHDwithoutcondition}
\end{equation}
Now consider the game payoff matrix $U$ so that payoffs obey $a+d=b+c$. We do so because unless this condition is obeyed, the partition function and expectation value for the order parameter (which we define later for finding the game magnetization) will not reduce to a concise analytical form in the thermodynamic limit of the game. In that case, numerical methods would be the only way to solve it. {This is infructuous since we aim to obtain an analytical formula to understand the game in thermodynamic limit}. When the payoff matrix obeys the condition $a+d=b+c$, eigenvalues of $K$ are $0$ and $(e^{\beta a}+e^{\beta d})$ and eigenvalues of $K^N$ are $0$ and $(e^{\beta a}+e^{\beta d})^N$. Since the sum of eigenvalues gives the trace of a matrix, we have
\begin{equation}
Z=Tr(K^N)=(e^{\beta a}+e^{\beta d})^N.
\label{partitionfunctionHD}
\end{equation}
To determine difference between number of players playing strategy $s_1$ and those playing strategy $s_2$, we introduce an operator $\hat{J_z}=\sum_iP_0^{(i)}-P_1^{(i)}$ where $i$ denotes player index\cite{Adami}. This operator is referred to as an order parameter. The expectation value of $\hat{J_z}$, which is game magnetization, gives, on average, the difference between the number of players playing $s_1$ strategy and players playing $s_2$ strategy in the thermodynamic limit. The magnetization per player is then given as
\begin{equation}
\langle \hat{J_z} \rangle_\beta=\frac{1}{N}\frac{1}{Z}\sum_{\ket{x}}\bra{x}\hat{J_z}e^{-\beta H}\ket{x}.
\end{equation}
For $N$ players, one can prove the following by mathematical induction,
\begin{align}
\sum_{\ket{x}}\bra{x}\hat{J_z}e^{-\beta H}\ket{x}=N(e^{\beta a}+e^{\beta d})^{N-1}(e^{\beta a}-e^{\beta d}).
\end{align}
Thus, one gets the average value of game magnetization per player as,
\begin{align}
\langle \hat{J_z} \rangle_\beta=&\frac{1}{N}\frac{\sum_{\ket{x}}\bra{x}\hat{J_z}e^{-\beta H}\ket{x}}{Z}=\frac{e^{\beta a}-e^{\beta d}}{e^{\beta a}+e^{\beta d}} \notag \\
=&\tanh\beta\bigg( \frac{a-d}{2}\bigg)
\label{magHD}
\end{align}
$\langle \hat{J_z} \rangle_\beta$ here is the HD model equivalent for $m_g$ of NE mapping.
%\textcolor{red}{The condition $a+d=b+c$ presents as a limitation only for the HD model. This condition was invoked in Eq.~(\ref{partitionfunctionHDwithoutcondition}) since the partition function and, henceforth, the game magnetization of the HD model will not converge to a concise mathematical formula when it is not obeyed. However, for game matrices that do not obey this condition, one can still solve it numerically through computation. We have not bothered with this in our work since we aimed to compare the analytical models with exact solutions.}
The game magnetization, as obtained by the HD model, fails to account for the Nash equilibrium points of the game in the thermodynamic limit, as we demonstrated through an analysis of the public goods game. It was first demonstrated in [Benjamin C, Sarkar S. The emergence of cooperation in the thermodynamic limit, Chaos Solitons Fractals 135,109762 (2020)] where the behaviour of game magnetization plotted against payoff was incompatible with the definition of the game being studied. Additionally, it was shown that the average payoff calculated at zero noise limit using the HD model did not correspond with the Nash equilibrium payoffs of the thermodynamic limit of the game.

\subsection{\label{DEtheory}Darwinian evolution (DE) model}
Ref.~\cite{Adami}, i.e., the paper that detailed the HD model, also introduced another method to derive the game magnetization, which produced slightly better results than the HD model, namely Darwinian evolution~\cite{Adami}. Here, we also follow the nomenclature of Ref.~\cite{Adami}. The name "Darwinian evolution" model is chosen because it refers to the case of a single player seeking the largest payoff without regard to the payoff received by other players in the system, akin to the "winner takes all" character of the Darwinian evolution process in nature. Darwinian evolution in nature does not optimize population fitness but maximizes a single individual's fitness within a population. DE model adopts this by giving each microstate of the system a probability weightage in the Boltzmann distribution with respect to the payoff of a single player as opposed to the combined payoff we see in the Hamiltonian dynamics model.

The analytical framework for the DE model is very similar to the HD model, except the Hamiltonian is chosen to focus on the energy of a single spin site or player. In this work, we choose the spin site or player in focus to be the one with index $i=1$. Starting from Hamiltonian in Eq.~(\ref{eqn:Ham}), we can see that we need to take into picture only two spin sites or two players, i.e., $i=1$ and its nearest neighbour $i=2$ to write the Hamiltonian corresponding to the energy of spin site $i=1$. It is given by
\begin{equation}
H'=H_1=\sum_{m,n=0}^1E_{mn}P_m^{(1)}\otimes P_n^{(2)}.
\end{equation}
While the HD model seeks to find collective minimum energy for all sites (or maximum payoffs for all players), the DE model minimizes the energy of a single focal spin site or maximizes the payoff of a single player. Hence magnetization derived for a given two-spin site (or two-player) system corresponds to minimizing energy (or maximizing payoff) of spin site $1$ when considering its interaction with spin site 2. The state of $2$-spin system is written as $\ket{x}=\ket{m_1m_2}$.

Partition function Z for two spin sites (or two players) is given by,
\begin{equation}
Z=\sum_{\ket{x}}\bra{x}e^{-\beta H_1}\ket{x}=\sum_{m_1m_2}\bra{m_1m_2}e^{-\beta H_1}\ket{m_1m_2}.
\end{equation}
As, $\bra{m_1m_2}H_1\ket{m_1m_2}=E_{m_1m_2}$,
\begin{equation}
\mbox {we get, } Z=\sum_{m_1m_2}e^{-\beta E_{m_1m_2}}.
\end{equation}
For sake of mathematical convenience, we introduce a matrix $K$ with elements defined as $K_{ij}=e^{-\beta E_{ij}}$, or $K=\begin{pmatrix} e^{\beta a} & e^{\beta b} \\
e^{\beta c} & e^{\beta d} \\
\end{pmatrix}$, via Eq.~(10). Thus, partition function $Z$ reduces to,
\begin{equation}
Z=\sum_{m_1m_2}K_{m_1m_2}=e^{\beta a} + e^{\beta b} + e^{\beta c} + e^{\beta d}.
\label{eqn:DEpartitionfunction}
\end{equation}
Unlike the HD model, where we require the payoff condition $a+d=b+c$ to get a simple analytical form for game magnetization, the DE model needs no such condition since the Hamiltonian is limited to a single spin site.
To determine difference between players playing strategy $s_1$ and those playing strategy $s_2$, we introduce an operator $\hat{J_z}=\sum_iP_0^{(i)}-P_1^{(i)}$ where $i$ denotes spin site index\cite{Adami}. The expectation value of $\hat{J_z}$ gives an average difference between several $s_1$ players and $s_2$ players in the thermodynamic limit. Since DE model only concerns itself with the game magnetization of focal player (or focal site) $1$, we have $\hat{J_z}^{(1)}=P_0^{(1)}-P_1^{(1)}$. Average game magnetization is then given as
\begin{equation}
\langle \hat{J_z}^{(1)} \rangle_\beta=\frac{1}{Z}\sum_{\ket{x}}\bra{x}\hat{J_z^{(1)}}e^{-\beta H_1}\ket{x}.
\end{equation}
Here we have $\ket{x}=\ket{m_1,m_2},m_i\in \{0,1\}$. Also we see that $\hat{J_z}^{(1)}\ket{0\hspace{0.5mm}m_2}=\ket{0\hspace{0.5mm}m_2}$ and $\hat{J_z}^{(1)}\ket{1\hspace{0.5mm}m_2}=-\ket{1\hspace{0.5mm}m_2}$. Thus,
\begin{align}
&\sum_{\ket{x}}\bra{x}\hat{J_z}^{(1)}e^{-\beta H_1}\ket{x}=\sum_{m_1m_2}\bra{m_1m_2}\hat{J_z}^{(1)}e^{-\beta H_1}\ket{m_1m_2} \notag\\
&=\sum_{m_2}e^{-\beta E_{0m_2}}-\sum_{m_2}e^{-\beta E_{1m_2}}=\sum_{m_2}K_{0m_2}-\sum_{m_2}K_{1m_2} \notag\\
&=e^{\beta a} + e^{\beta b} - e^{\beta c} - e^{\beta d},
\end{align}
wherein, $K$ represents matrix in Eq.~(\ref{K}).

We thus have average magnetization for the focal player as,
\begin{align}
\langle \hat{J_z}^{(1)} \rangle_\beta=&\frac{1}{Z}\sum_{\ket{x}}\bra{x}\hat{J_z}^{(1)}e^{-\beta H}\ket{x}, \notag \\
=&\frac{e^{\beta a} + e^{\beta b} -
e^{\beta c} - e^{\beta d}}{e^{\beta a} + e^{\beta b} +
e^{\beta c} + e^{\beta d}},
\label{eqn:MagDEgeneral}
\end{align}
in general. In case payoffs obey condition $a+d=b+c$, we get average magnetization as,
\begin{align}
\langle \hat{J_z}^{(1)} \rangle_\beta=&\frac{(e^{\beta a} -
e^{\beta c})(1+e^{\beta (b-a)})}{(e^{\beta a}+
e^{\beta c})(1+e^{\beta (b-a)})}=\frac{(e^{\beta a} -
e^{\beta c})}{(e^{\beta a}+
e^{\beta c})} \notag \\
=&\tanh\beta\bigg(\frac{a-c}{2}\bigg).
\label{eqn:MagDE}
\end{align}
$\langle \hat{J_z}^{(1)} \rangle_\beta$ here is DE model equivalent for $m_g$ of NE mapping. For condition $a+d=b+c$, game magnetization given by NE mapping and DE model is identical. However, when $a+d\neq b+c$, game magnetization obtained from DE model and NE mapping differ. Game magnetization from the HD model differs from both NE mapping and the DE model. The nomenclature "Darwinian evolution" model can be justified as follows: the idea is to represent the dynamics of a system where a single player seeks the largest payoff without regard to the payoff received by other players in the system (akin to Darwinian evolution in nature where an individual's survivability depends on how they can increase their fitness while disregarding that of others).
\subsection{\label{ABMtheory}Agent based method (ABM)}
A popular numerical approach to simulate the thermodynamic limit of social dilemma games is via Agent based method. For this simulation, we consider $10,000$ players/sites. These 10,000 players are arranged in a $1D$ chain, each interacting with its nearest neighbour. Each site's energy is taken per energy matrix $E$, obtained by taking the negative of the game payoff matrix as in Eq.~(\ref{payoffandenergymatrix}). We update players' strategy $10$ million times for a randomly chosen player in the $1D$ chain, i.e., on an average of $10000$ updates per player. The algorithm is summed up as follows:
\begin{enumerate}
\item Assign random strategy to all players, i.e., $(Hawk/dove)$ or $(provide/free-ride)$ depending on the game.
\item Randomly chooses a focal player and finds the strategy assigned to it and its nearest neighbour. Depending on these strategies, its energy $E$ is determined.
\item We determine the difference in energy $\Delta E_i$ if the focal player had chosen another strategy while keeping the strategy of its nearest neighbour fixed.
\item Flip strategy of focal player [Hawk/Provide to Dove/free-ride or vice versa] according to probability given by the Fermi function $1/(1+e^{\beta \Delta E_i})$\cite{altrock,bladon}.
\item Go to process 2 (repeat, say, 10000 times).
\item Calculate the difference between the fraction of cooperators and defectors.
\end{enumerate}
We can see that probability of flipping increases when $\Delta E_i$ decreases, i.e., the lesser the energy difference, the greater the probability of flipping. The reason for minimizing energy is for the system to reach equilibrium at the lowest energy state. Historically, mathematicians were concerned with dynamical processes that could explain the distribution of strategies among a population. For example, Ref.~\cite{altrock} studies evolutionary dynamics for well-mixed populations (where every player interacts with every other player) and how it converges to homogeneous states (where all players assume the same strategy) using recursion equations. Ref.~\cite{bladon} approaches evolutionary dynamics again for a well-mixed population using replicator dynamics modified to allow for mutations and additional conversion rates of strategy happening over time due to reproduction and death occurring at a rate dependent on reproductive fitness. Comparatively, our model is much simpler since we are not interested in dynamical processes; instead, we use equilibrium statistical mechanics to study a system where interaction between players is limited to between nearest neighbours in a $1D$ chain. However, we will be using the Fermi function for our Agent-based simulation, which is common to both of these papers, which they use to show microscopic transition rates in strategies by the players. The Python 3 code for the Agent-based simulation of the Hawk-Dove and Public Goods game is given in Appendix A. In the following two sections, we will utilize each analytical model to understand the social dilemmas in the thermodynamic limit and compare them to the Agent-based method.
\section{Hawk-Dove game}
The Hawk-Dove game is a $2$-player $2$-strategy game of conflict wherein the two strategies either yield or risk some damage. It is usually explained in the context of a situation when two individuals fight over a common resource ($r$). When both players choose to be Hawk, they suffer injury denoted by payoff $-s$, with $s>0$. When one goes for Hawk and the other for Dove, the Hawk player gains resource indicated by payoff $r$ while the Dove player loses it indicated by payoff $-r$, with $r>0$. When both players adopt the Dove strategy, they earn nothing indicated by zero payoffs. Note that $s>r>0$ since injury caused by fighting outweighs the gained resource. The payoff matrix for the game is given by,
\begin{equation}
U=
\left(\begin{array}{c|c c}
& Hawk & Dove \\
\hline
Hawk & -s,-s & r,-r \\
Dove & -r,r & 0,0
\end{array}\right)
\label{chickengamematrix}.
\end{equation}
This game has two pure strategy Nash equilibria— $(Dove, Hawk)$ and $(Hawk, Dove)$, and a mixed strategy Nash equilibrium $(\sigma,\sigma)$ where $\sigma$ is the state where the player chooses to be Hawk with probability $p$ and Dove with probability $(1-p)$, with $p=r/s$ \cite{schecter}. Note that this game does not obey condition $a+d=b+c$. Hence we cannot use the HD model to study this game. Therefore we compare NE mapping with DE model and Agent-based method.
\subsection{\label{Chicken-NE}Results from NE mapping}
This section will analyze the game magnetization and the average payoff per player using NE mapping.
\subsubsection{\label{Chicken-NE-mag}Game magnetisation}
For the Hawk-Dove game, we get the NE mapping game magnetization from Eqs.~(\ref{eqn:magNEmappingabcdgeneral},\ref{chickengamematrix}) as,
\begin{equation}
m_g=\frac{\sinh \beta(\frac{2r-s}{4})}{\sqrt{\sinh^2 \beta(\frac{2r-s}{4})+e^{\beta s}}}.
\label{eqn:NEmagchicken}
\end{equation}
Most players choose the $Hawk$ strategy when $r > s/2$, while most choose the $Dove$ strategy when $r<s/2$. Notably, we see game magnetization being zero at $r = s/2$. It is because NE mapping picks out mixed strategy Nash equilibrium $(\sigma,\sigma)$ in this case and follows that path. At $r = s/2$, Nash equilibrium is where players have an equal probability of choosing either Hawk or Dove strategy, and game magnetization become zero. Game magnetization vanishes in infinite noise (i.e., $\beta \xrightarrow{} 0$ limit. It is because noise randomizes each player's strategy. At infinite noise, players choose strategies randomly, leading to an equal number of Hawks and Doves, giving game magnetization as zero. Interestingly, we see game magnetization being zero in zero noise (i.e., $\beta \xrightarrow{} \infty$ ) limit.
\subsubsection{\label{Chicken-NE-avgU}Average payoff per player}
We start with writing a transformed payoff matrix for the Hawk-Dove game where transformations are done according to Eq.~(\ref{eqn:transformedpayoffmatrix}),
\begin{equation}
U'=
\left(\begin{array}{c c}
\frac{r-s}{2},\frac{r-s}{2} & \frac{r}{2},-\frac{s-r}{2} \\
\frac{s-r}{2},\frac{r}{2} & \frac{-r}{2},\frac{-r}{2}
\end{array}\right).
\label{transformedChickenmatrix}
\end{equation}
In order to find the average payoff per player, we start with Hamiltonian of $2$-site $1D$ Ising system given as,
\begin{equation}
H=-J(\sigma_1\sigma_2+\sigma_2\sigma_1)-h(\sigma_1+\sigma_2),
\label{H-2site}
\end{equation}
where $\sigma_i, i=1,2$ denote up and down spins. The partition function of this $2$-site Ising system is then given as,
\begin{equation}
Z=\sum_{\sigma_1,\sigma_2}e^{-\beta H}=e^{\beta(2J+2h)}+2e^{-\beta(2J)}+e^{\beta(2J-2h)},
\label{H-2site,Z-2site}
\end{equation}
where $\beta=1/(k_BT)$. Average thermodynamic energy $\langle E \rangle$ can be obtained\cite{Reif} from partition function as,
\begin{equation}
\langle E \rangle=-\frac{\partial \ln Z}{\partial \beta}.
\label{eqn:averagethermodynamicenergy}
\end{equation}
Hence, we get average thermodynamic energy in terms of Ising parameters $J$ and $h$ as,
\begin{equation}
\langle E \rangle=-\frac{\resizebox{.4 \textwidth}{!} {$(2J+2h)e^{\beta (2J+2h)}-2(2J)e^{-\beta (2J)}+(2J-2h)e^{\beta (2J-2h)}$}}{e^{\beta (2J+2h)}+2e^{-\beta (2J)}+e^{\beta (2J-2h)}}.
\label{avgthermodynamicenergyJandh}
\end{equation}
In order to find the average payoff, we take the negative of average thermodynamic energy since we equate payoffs with negative energy. Further, we divide this by $2$ to get the average payoff per player $\langle U' \rangle$ (since we took the energy of $2$ to spin sites for the partition function in the beginning). We substitute $J$ and $h$ in terms of game payoffs from Eq.~(\ref{eqn:Jandh}) as follows- $J=\frac{a+d-b-c}{4}=\frac{-s+r-r}{4}=s/4$ and $h=\frac{a+b-c-d}{4}=(2r-s)/4$. Hence, we get the average payoff per player as,
\begin{align}
\langle U' \rangle= -\frac{1}{2}\langle E \rangle=\frac{\frac{r-s}{2}e^{\beta(r-s)}+\frac{s}{2}e^{\beta(s/2)}-\frac{r}{2}e^{-\beta r}}{e^{\beta(r-s)}+2e^{\beta(s/2)}+e^{-\beta r}}.
\label{Chickengameaveragepayoff}
\end{align}
In $\beta \xrightarrow{} 0$ limit, i.e., increasing noise in the system leading to randomness in choice of strategies, we get the average payoff per player from Eq.~(\ref{Chickengameaveragepayoff}) as,
\begin{equation}
\underset{\beta \xrightarrow{} 0}{\lim} \langle U' \rangle = 0.
\end{equation}
The average payoff is zero in the $\beta \xrightarrow{} 0$ limit. From Eq.~(\ref{transformedChickenmatrix}), this is nothing but the average of payoffs of \textit{(Hawk, Hawk), (Hawk, Dove), (Dove, Hawk)} and \textit{(Dove, Dove)} strategies: $\frac{(r-s)/2+r/2-r/2+(s-r)/2}{4}=0$. It happens because, at $\beta \xrightarrow{} 0$, all strategy pairs become equiprobable due to complete randomization in the players' choice of strategy. In terms of original game payoffs (see Eq.~(\ref{chickengamematrix})), an average of all strategy pairs gives,
\begin{equation}
\underset{\beta \xrightarrow{} 0}{\lim} \langle U \rangle = -s/4.
\end{equation}
Now we proceed to find the average payoff per player in zero noise, i.e., no randomness in the choice of strategy ( $\beta \xrightarrow{} \infty$), limit. We do this because noise inherently randomizes each player's strategy, and there is no randomization for zero noise. Calculating the average payoff per player in zero noise limit and comparing it to Nash equilibrium payoff of $2$-player $2$-strategy game, we will get an idea about how accurately NE mapping predicts the equilibrium behaviour of players for zero noise. Since, payoffs $s>r>0$, taking zero noise (or, $\beta \xrightarrow{} \infty$) limit in Eq.~(\ref{Chickengameaveragepayoff}), gives
\begin{equation}
\underset{\beta \xrightarrow{} \infty}{\lim} \langle U' \rangle=s/4.
\end{equation}
Examining this, we can see that the average payoff per player in zero noise (i.e., $\beta \xrightarrow{} \infty$) limit is average of payoffs for strategies $(Hawk, Dove)$ and $(Dove, Hawk)$ respectively in the transformed payoff matrix $U'$ of Eq.~(\ref{transformedChickenmatrix}), i.e., $\frac{r/2+(s-r)/2}{2}=\frac{s}{4}$. These strategy pairs are pure strategy Nash equilibrium for the Hawk-Dove game. The average payoff for strategies $(Hawk, Dove)$ and $(Dove, Hawk)$ in terms of original game payoffs (see, Eq.~(\ref{chickengamematrix})), is
\begin{equation}
\underset{\beta \xrightarrow{} \infty}{\lim} \langle U \rangle=0.
\end{equation}
It gives us a picture of the Hawk-Dove game in thermodynamic limit wherein players alternate between $Hawk$ and $Dove$ strategies. It is also consistent with game magnetization in zero noise (or, $\beta \xrightarrow{} \infty$) limit vanishing due to half the number of players choosing $Dove$ while the other half chooses $Hawk$.
\subsection{\label{Chicken-DE}Results from DE model}
In this section, we will analyze game magnetization for the Hawk-Dove game and the average payoff per player using the DE model.
\subsubsection{Game magnetisation}
We have $a=-s, b=r,c=-r$ and $d=0$ for the Hawk-Dove game. We get game magnetization using the DE model from Eq.~(\ref{eqn:MagDEgeneral}) as,
\begin{equation}
\langle \hat{J_z}^{(1)}\rangle_\beta=\frac{e^{-\beta s}+e^{\beta r}-e^{-\beta r}-1}{e^{-\beta s}+e^{\beta r}+e^{-\beta r}+1}.
\label{MagDEChicken}
\end{equation}
The DE model's game magnetization does not match that derived from NE mapping. The reasons for this will be analyzed in Sec.~\ref{Chicken-Analysis}. Game magnetization vanishes in infinite noise (i.e., $\beta \xrightarrow{} 0$ limit indicating an equal number of Hawk and Dove players. In the zero noise (i.e., $\beta \xrightarrow{} 0$ limit, game magnetization is $1$, indicating all players chose the $Hawk$ strategy.
\subsubsection{Average payoff per player}
In order to determine the average payoff per player, we consider the partition function of the DE model in Eq.~(\ref{eqn:DEpartitionfunction}) for the Hawk-Dove game, which is,
\begin{equation}
Z=e^{-\beta s}+e^{\beta r}+e^{-\beta r}+1.
\end{equation}
As we did before, we find average thermodynamic energy $\langle E \rangle$ using Eq.~(\ref{eqn:averagethermodynamicenergy}) and then take its negative to obtain average payoff per player $\langle U \rangle$. We do not divide by $2$ since the Hamiltonian for DE model is concerned with the energy of a single spin site, thus,
\begin{equation}
\langle U \rangle=-\langle E \rangle=\frac{\partial \ln Z}{\partial \beta}
=\frac{-se^{-\beta s}+re^{\beta r}-re^{-\beta r}}{e^{-\beta s}+e^{\beta r}+e^{-\beta r}+1}.
\end{equation}
In infinite noise (i.e., $\beta \xrightarrow{}0$ limit, we get average payoff per player as,
\begin{equation}
\underset{\beta \xrightarrow{} 0}{\lim} \langle U \rangle = -s/4.
\end{equation}
As we saw before, this happens because, at infinite noise, all strategy pairs become equiprobable. Hence, we get $\langle U \rangle$ as the average of four payoffs in the payoff matrix (\ref{chickengamematrix}). Since $s>r>0$, taking zero noise (i.e., $\beta \xrightarrow{} \infty$ limit, we get average payoff per player as
\begin{equation}
\underset{\beta \xrightarrow{} \infty}{\lim} \langle U \rangle=r.
\end{equation}
It is incorrect for a $1D$ chain of players where the nearest neighbours are engaged in the Hawk-Dove game, as there is no way every player can get an average payoff of $r$. Additionally, we see game magnetization being $1$ in zero noise limit indicating all players choose the $Hawk$ strategy, which should give us the average payoff per player as $s$ instead. The average payoff here is the largest in the payoff matrix and does not match the game's Nash equilibrium. Hence, the DE model fails to give plausible results for the Hawk-Dove game in the zero noise limit.
\subsection{\label{Chicken-ABM}Results from agent based method}
In this section, we analyze game magnetization and average payoff per player using a numerical Agent-based method.
\subsubsection{\label{Chicken-ABM-Mag}Game magnetisation}
Here we take energy matrix $E$ as negative of the payoff matrix in Eq.~(\ref{chickengamematrix}) and proceed with the algorithm as described in Sec.~\ref{ABMtheory} to find game magnetization,
\begin{equation}
\text{thus, }E=
\bigg(\begin{array}{c c}
s &-r\\
r & 0
\end{array}\bigg).
\label{eqn:ChickenAgentenergymatrix}
\end{equation}
In this energy matrix, we have only written payoffs of row players since it is a symmetric game. Game magnetization obtained via Agent-based simulation is shown in Fig.~\ref{fig:MagChicken}, and its analysis is done in Sec.~\ref{Chicken-Analysis-Mag}.
\subsubsection{\label{Chicken-ABM-avgU}Average payoff per player}
Using Agent-based simulation, we find the average payoff per player with zero noise limit. Here, we follow the same algorithm as described in Sec.~\ref{ABMtheory}, and the simulation is carried out at very low noise levels (i.e., very high selection intensity, $\beta=10^6$), in the thermodynamic limit of Hawk-Dove game. We sum the energies of all spin sites and divide them by the total number of spin sites to get the average energy per site. Since we took the energy matrix as a negative of the game payoff matrix, we took the negative of this value to get the average payoff per player. The average payoff per player in zero noise ($\beta \rightarrow \infty$) limit obtained via Agent-based simulation is shown in Fig.~\ref{fig:ChickengameaverageUcombined}, and analysis is done in Sec.~\ref{Chicken-Analysis-avgU}. For infinite noise (i.e., $\beta \xrightarrow{} 0$ limit), game magnetization vanishes, and the average payoff per player equals $-s/4$. In zero noise (i.e., $\beta \xrightarrow{} \infty$ limit), we find game magnetization and average payoff per player both vanishing.

\subsection{\label{Chicken-Analysis}Analysis of Hawk-Dove game}
Herein we discuss game magnetization as well as average payoff per player calculated in Sec.~\ref{Chicken-NE} for NE mapping, Sec.~\ref{Chicken-DE} for DE model, and Sec.~\ref{Chicken-ABM} for Agent-based method for Hawk-Dove game. We especially focus on anomalies in the DE model and try to understand their reasons.
\subsubsection{\label{Chicken-Analysis-Mag}Game magnetisation}
\begin{figure}
\centering
\includegraphics[width=8.5cm]{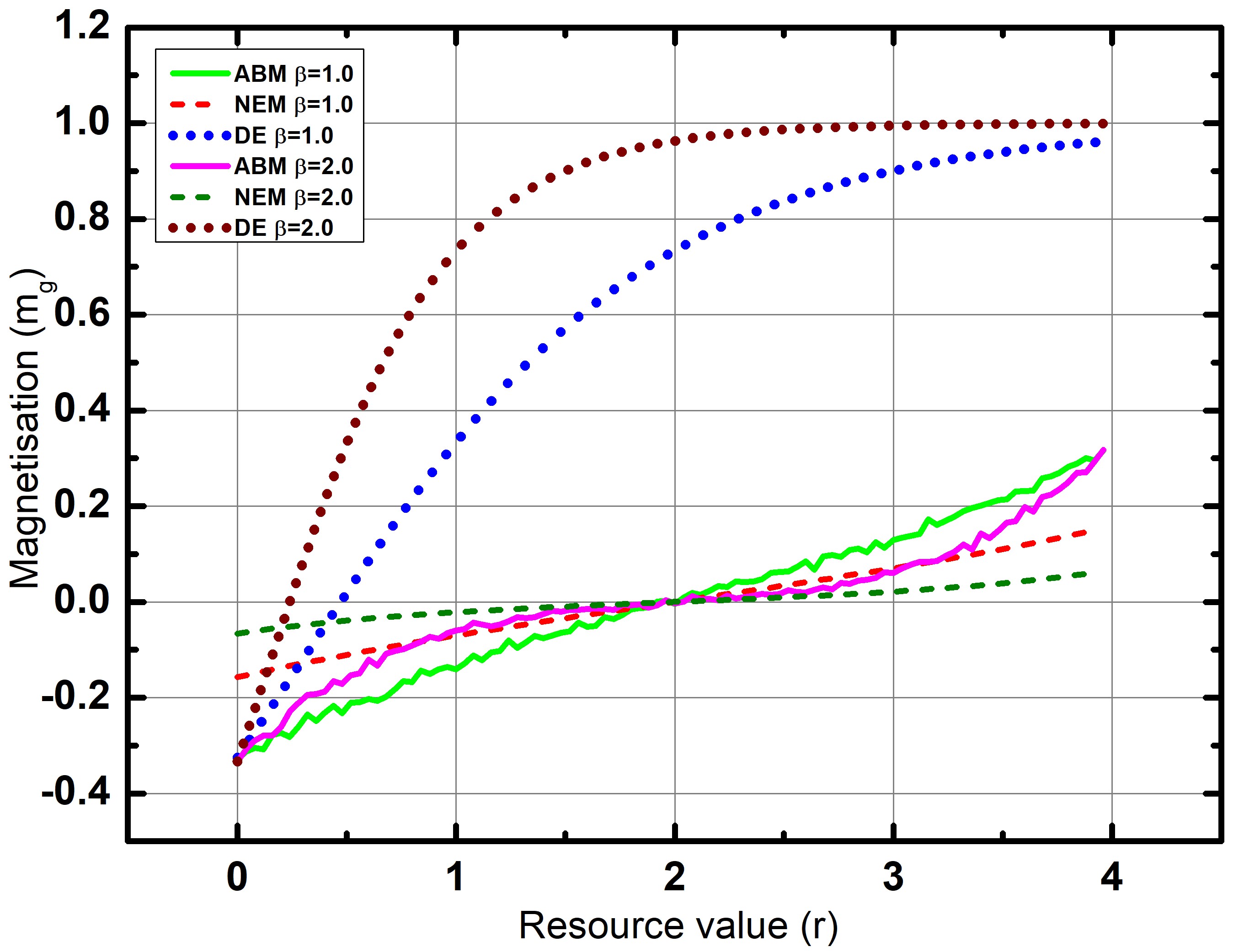}
\caption{Game magnetization $m_g$ vs. resource value $r$ for the Hawk-Dove game as obtained by NE mapping (NEM), Darwinian evolution (DE) and agent-based method (ABM) for the cost of injury $s=4$ and for $\beta=1.0$ and $\beta=2.0$.}
\label{fig:MagChicken}
\end{figure}
Using NE mapping and DE models, we have analytically obtained game magnetization and average payoff per player for the Hawk-Dove game. We also looked at zero noise (high selection intensity, i.e., $\beta \xrightarrow{} \infty$) and infinite noise (zero selection intensity, i.e., $\beta \xrightarrow{} 0$) limits for game magnetization and average payoff per player respectively. We compare this with numerical results obtained via Agent based method. This subsection will examine game magnetization vs. resource value for the two analytical models and the Agent-based method.\\

In Fig.~\ref{fig:MagChicken}, we have plotted game magnetization against resource value as obtained by NE mapping (dashed red and green lines), DE model (dotted blue and dark red lines), and agent-based simulation (solid black and pink lines), at $\beta=1.0$, and $\beta=2.0$ with the cost of injury fixed at $s=4$. For NE mapping, we see a phase transition at $r = s/2$. At $r = s/2$, players divide equally between Hawk and Dove strategies. For $r > s/2$, more players choose Hawk than Dove; hence, game magnetization is greater than zero. For $r < s/2$, more players choose Dove than Hawk; hence, game magnetization is less than zero. Further, game magnetization for NE mapping is in close agreement with that obtained from the Agent-based method, and the two curves intersect precisely at the same transition point. Conversely, the game magnetization plot obtained via the DE model deviates significantly from both NE mapping and the Agent-based method. Additionally, game magnetization obtained via the DE model does not indicate the Nash equilibria of the game. Finally, while NE mapping and agent-based simulation pick out the phase transition point at the same resource value $r = s/2$, the DE model predicts the phase transition around $r\xrightarrow{}0$, which is incorrect.
\subsubsection{\label{Chicken-Analysis-avgU}Average payoff per player}
\begin{figure}
\includegraphics[width=8.5cm]{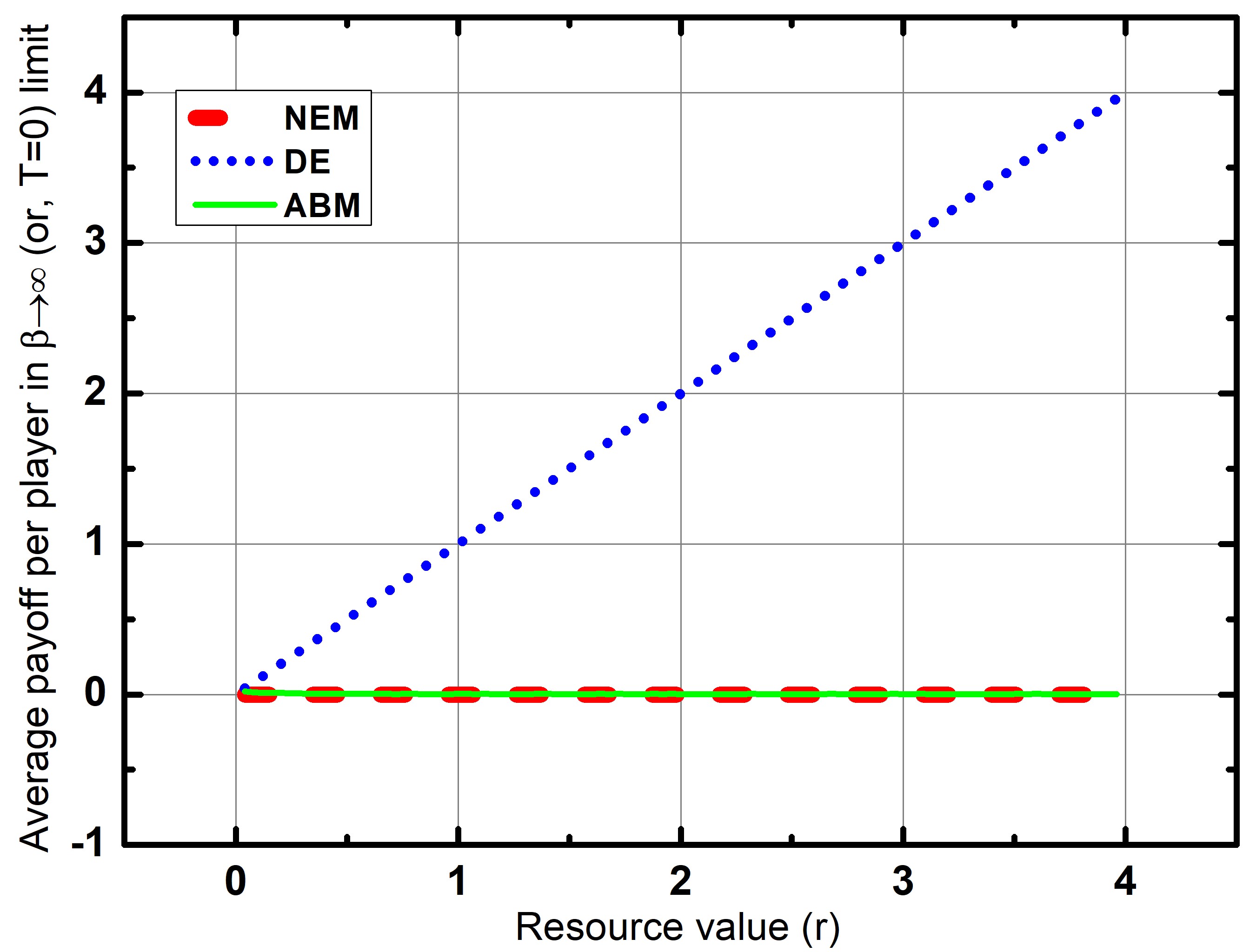}
\caption{$\underset{\beta \xrightarrow{} \infty}{lim} \langle U \rangle$ vs. $r$ for the Hawk-Dove game as obtained by NE mapping (NEM), Darwinian evolution (DE) model and agent-based method (ABM) for punishment $p=1$ and cost $t=6$.}
\label{fig:ChickengameaverageUcombined}
\end{figure}
In Fig.~\ref{fig:ChickengameaverageUcombined}, we plot the average payoff per player in zero noise (or, $\beta \xrightarrow{} \infty$) limit vs. resource value as obtained via NE mapping (dashed red line), DE model (dotted blue line) and Agent-based method (solid black line) for Hawk-Dove game with the cost of injury $s=4$. We see that the average payoff per player for NE mapping is zero regardless of the resource value $r$. It is because half of the players adopt Hawk while the other half adopt Dove. Since NE mapping picks out pure strategy equilibria- $(Hawk,Dove)$ and $(Dove,Hawk)$- in the zero noise (or, $\beta \xrightarrow{} \infty$) limit. In such a state, players get payoffs either $r$ or $-r$, giving the average payoff per player in zero noise ($\beta \xrightarrow{} \infty$) limit as zero. For the DE model, the average payoff per player in zero noise limit equals $r$, which is only possible when each player gets a payoff $r$ (since $r$ is the largest payoff in the payoff matrix). However, it is impossible since the only way a player gets a payoff $r$ is when the player adopts strategy $Hawk$. Its interacting neighbour adopts strategy $Dove$- implying that both neighbouring players cannot get the same payoff $r$. Hence, the DE model fails to give a consistent picture of interacting players in the Hawk-Dove game. For Agent based method, we get the average payoff per player in zero noise limit as zero regardless of the resource value $r$. Thus, NE mapping perfectly agrees with results from Agent-based simulation in the zero noise (or, $\beta \xrightarrow{} \infty$) limit of the Hawk-Dove game.

\section{Public goods game}
Public goods game is a social dilemma game where individuals choose to either contribute (provide) or not (free-ride) a certain amount of tokens (or cost $t$) to be put into a public pot. The net amount of tokens in the pot is multiplied by a factor $k>1$, i.e., the net contribution from players is used to generate positive interest, and revenue is divided among all players in the group. Payoffs for an individual when they cooperate or provide ($P_C$) or defect or free-ride ($P_D$) is given \cite{Colin2} by,
\begin{equation}
P_D=\frac{kN_Ct}{N}\hspace{5mm} and \hspace{5mm}P_C=P_D-t.
\label{P_CP_D}
\end{equation}
In Eq.~(\ref{P_CP_D}), $N_C$ is the number of players who cooperate, $t$ is the cost or amount of tokens each player contributes if they choose to, and $N$ is the total number of players. We study public goods game involving two players, i.e., $N=2$. When both players co-operate ($N_C=2$) they get payoff $(kt-t),$ with $k>1$. When one defects and other cooperates ($N_C=1$), defecting player gets payoff $\frac{kt}{2}$ and providing player gets $(\frac{kt}{2}-t)$. When both players provide, both get zero payoffs. By substituting for payoff $(kt-t)=2r$, we get $(\frac{kt}{2}-t)=(r-\frac{t}{2})$ and $\frac{kt}{2}=(r+\frac{t}{2})$, $r$ denotes reward. In a more general case, one can also punish players who choose to free-ride to reduce the tendency for individuals to always free-ride. It is done by adding a negative payoff $-p$ (with $p>0$) to the payoff of a player who chooses to free-ride; $ p$ is called punishment. The payoff matrix for such a game is written as
\begin{equation}
U=
\left(\begin{array}{c|c c}
& provide & free-ride \\
\hline
provide & 2r,2r & r-\frac{t}{2},r+\frac{t}{2}-p \\
free-ride & r+\frac{t}{2}-p,r-\frac{t}{2} & -p,-p
\end{array}\right).
\label{eqn:publicgoodsgamematrix}
\end{equation}
One can see both the public goods game with punishment (\ref{eqn:publicgoodsgamematrix}) or that without punishment ($p=0$) obey payoff condition $a+d=b+c$, unlike the Hawk-Dove game. We can see that for $r > (\frac{t}{2}-p)$, Nash equilibrium is $(provide,provide)$ (i.e., $(cooperate,cooperate)$) while for $r < (\frac{t}{2}-p)$, Nash equilibrium is $(free-ride,free-ride)$, i.e., $(defect,defect)$. Now we will analyze game magnetization and average payoff per player in both zero noise and infinite noise limits using the HD model, DE model, NE mapping, and Agent-based method.
\subsection{\label{PG-NE}Results from NE mapping}
This section will analyze game magnetization and average payoff per player using NE mapping.
\subsubsection{\label{PG-NE-mag}Game magnetisation}
For the public goods game, comparing elements of payoff matrices in Eq.~(\ref{eqn:publicgoodsgamematrix}) and Eq.~(\ref{eqn:abcdmatrix}), we have $a=2r, b=r-\frac{t}{2},c=r+\frac{t}{2}-p,$ and $d=-p$. Since payoffs obey condition $a+d=b+c$, we have game magnetisation obtained from Eq.~(\ref{magNEmappingabcdwithcondition}) as
\begin{equation}
m_g=\tanh\beta\bigg(\frac{2r-t+2p}{4}\bigg).
\label{eqn:NEmagPublicgoods}
\end{equation}
In the infinite noise or zero selection intensity ($\beta \xrightarrow{} 0$ limit, we see that game magnetization vanishes due to the complete randomization of strategies leading to equal numbers of providers and free riders. In zero noise or high selection intensity ($\beta \xrightarrow{} \infty$) limit, we get game magnetisation as $+1$ (all provide) when $r>(t/2-p)$ and $-1$ (all free-ride) when $r < (t/2-p)$. Hence, for public goods games, NE mapping drives players towards the Nash equilibrium strategy in the thermodynamic limit in zero noise limit.
\subsubsection{\label{PG-NE-avgU}Average payoff per player}
Since NE mapping uses a transformed payoff matrix in Eq.~(\ref{eqn:transformedpayoffmatrix}) to map the $2$-site Ising model to the payoff matrix of the game, we will see how the transformed payoff matrix will look like for the public goods game,
\begin{equation}
U'=
\left(\begin{array}{c c}
\frac{2r-t+2p}{4},\frac{2r-t+2p}{4} & \frac{2r-t+2p}{4},-\frac{2r-t+2p}{4} \\
-\frac{2r-t+2p}{4},\frac{2r-t+2p}{4} & -\frac{2r-t+2p}{4},-\frac{2r-t+2p}{4}
\end{array}\right).
\label{PGtransformedmatrix}
\end{equation}
To derive the average payoff per player for public goods game, we start with the partition function for $2$-site $1D$ Ising model, see Eq.~(\ref{H-2site,Z-2site}),
\begin{equation}
Z=e^{\beta(2J+2h)}+2e^{-\beta(2J)}+e^{\beta(2J-2h)}.
\end{equation}
We get average thermodynamic energy using Eq.~(\ref{avgthermodynamicenergyJandh}) in terms $J$ and $h$ as,
\begin{align}
\langle E \rangle&=-\frac{\partial \ln Z}{\partial \beta} \notag \\
&=-\frac{\resizebox{.4 \textwidth}{!} {$(2J+2h)e^{\beta (2J+2h)}-2(2J)e^{-\beta (2J)}+(2J-2h)e^{\beta (2J-2h)}$}}{e^{\beta (2J+2h)}+2e^{-\beta (2J)}+e^{\beta (2J-2h)}}.
\end{align}
To find the average payoff per player $\langle U' \rangle$, we take the negative of half of the average thermodynamic energy as we did in Sec.~\ref{Chicken-NE-avgU}. We substitute $J$ and $h$ in terms of game payoffs from Eq.~(\ref{eqn:Jandh}), i.e., $J=0$ and $h=(2r-t+2p)/4$. Hence, we have an average payoff per player as
\begin{align}
&\langle U' \rangle=-\frac{1}{2}\langle E \rangle=
\frac{\resizebox{.3 \textwidth}{!} {$\frac{2r-t+2p}{4}e^{\beta(\frac{2r-t+2p}{2})}-\frac{2r-t+2p}{4}e^{-\beta(\frac{2r-t+2p}{2})}$}}{e^{\beta(\frac{2r-t+2p}{2})}+e^{-\beta(\frac{2r-t+2p}{2})}}.
\label{publicgoodsaverageU}
\end{align}
In infinite noise (or, $\beta \xrightarrow{} 0$) limit, we get the average payoff per player as,
\begin{equation}
\underset{\beta \xrightarrow{} 0}{\lim} \langle U' \rangle = 0.
\end{equation}
Again, this is because all strategy pairs have the same probability of being chosen by interacting players. However, $\langle U' \rangle $ denotes the average payoff from the transformed payoff matrix. When we write this in terms of untransformed public goods game payoffs, see (\ref{eqn:publicgoodsgamematrix}), we get,
\begin{equation}
\underset{\beta \xrightarrow{} 0}{\lim} \langle U \rangle = r - p/2.
\end{equation}
In zero noise or high selection intensity (i.e., $\beta \xrightarrow{} \infty$) limit, we see that average payoff per player $\langle U' \rangle$, from Eq.~(\ref{publicgoodsaverageU}) depends on the sign of $2r-t+2p$.
\begin{enumerate}
\item For $(2r-t+2p)>0$, in high selection intensity (or, $\beta\xrightarrow{}\infty$) limit, $e^{-\beta (2r-t+2p)/2}$ goes to zero giving us average payoff per player as,
\begin{equation}
\underset{\beta \xrightarrow{} \infty}{\lim}\langle U' \rangle=\frac{2r-t+2p}{4}.
\end{equation}
\item For $(2r-t+2p)<0$, in high selection intensity (or, $\beta\xrightarrow{}\infty$) limit, $e^{\beta (2r-t+2p)/2}$ goes to zero giving us average payoff per player as,
\begin{equation}
\underset{\beta \xrightarrow{} \infty}{\lim}\langle U' \rangle=-\frac{2r-t+2p}{4}.
\end{equation}
\end{enumerate}
Thus, for $r > (t/2-p)$, we get the average payoff per player in infinite noise (or, $\beta \xrightarrow{} \infty$) limit as $(2r-t+2p)/4$. It corresponds to the payoff of strategy pair $(provide, provide)$ in transformed payoff matrix $U'$ of Eq.~(\ref{PGtransformedmatrix}), which is also Nash equilibrium for $r > (t/2-p)$. Similarly, for $r<(t/2-p)$, we have an average payoff per player in zero noise or high selection intensity (i.e., $\beta \xrightarrow{} \infty$) limit as $-(2r-t+2p)/4$. It corresponds to the payoff for strategy pair $(free-ride,free-ride)$ in transformed payoff matrix $U'$, which is Nash equilibrium for $r < (t/2-p)$. Thus, we can write these limits in terms of the untransformed public goods game payoffs as,
\begin{equation}
\underset{\beta \xrightarrow{} \infty}{\lim}\langle U \rangle=
\begin{cases}
2r & , r > (t/2-p). \\
-p & , r < (t/2-p).
\end{cases}
\end{equation}
Therefore, without randomizing strategies, NE mapping always gives payoffs corresponding to the Nash equilibrium of the public goods game. We will further analyze these results in Sec.~\ref{PG-Analysis}.
\subsection{\label{PG-HD}Results from HD model}
This section will analyze game magnetization and average payoff per player using an HD model.
\subsubsection{\label{PG-HD-Mag}Game magnetization}
For public goods game with punishment, we have $a=2r,b=r-t/2,c=r+t/2-p$, and $d=-p$. From Eq.~(\ref{magHD}) we have game magnetization obtained via HD model as,
\begin{equation}
\langle\hat{J_z}\rangle_\beta = \tanh \beta \bigg( \frac{2r+p}{2}\bigg).
\end{equation}
Thus, for $r > 0$ and $p > 0$, game magnetization is always positive, i.e., the majority of players choose to provide even when Nash equilibrium strategy is defection in regime $r < (t/2-p)$. Also, game magnetization remains independent of cost $t$. As expected, game magnetization vanishes for infinite noise, zero selection intensity (or, $\beta \xrightarrow{} 0$) limit. It is because, for increasing noise (low selection intensity), randomization in the choice of strategy of players increases. When noise tends to infinity (i.e., zero selection intensity $\beta \xrightarrow{} 0$), the choice of strategy is completely randomized, and players end up being cooperators and defectors in more or less equal proportion, hence overall, game magnetization vanishes. In zero noise (high selection intensity or, $\beta \xrightarrow{} \infty$) limit, we see that game magnetization equals $+1$, since $(2r+p)$ is always greater than zero, for public goods game, implying all players provide. It is because the HD model seeks to maximize the collective payoff of all players. In zero noise ($\beta \xrightarrow \infty$) limit, randomization in the choice of strategy vanishes. Since the collective payoff is largest when the interacting neighbours provide, all players provide with absolute certainty. In the analysis section, we will focus on the results and how they compare to results from other models.
\subsubsection{\label{PG-HD-avgU}Average payoff per player}
Now we seek to find the average payoff per player for the HD model, and we start with the partition function, from Eq.~(\ref{partitionfunctionHD}), written in terms of payoffs (for $N=2$ players) as,
\begin{equation}
Z=(e^{\beta 2r}+e^{-\beta p})^2.
\end{equation}
Average payoff per player $\langle U \rangle$ is again negative of average thermodynamic energy $\langle E \rangle$ and since energy matrix is taken as negative of payoff matrix and partition function is written for a system with $2$ players, therefore to calculate $\langle U \rangle$ we divide by the number of players. Thus,
\begin{equation}
\langle U \rangle=-\frac{1}{2}\langle E \rangle=\frac{1}{2}\frac{\partial \ln Z}{\partial \beta}=
\frac{2re^{\beta (2r)}-pe^{-\beta p}}{e^{\beta (2r)}+e^{-\beta p}}.
\label{eqn:HDavgU}
\end{equation}
In infinite noise ($\beta \xrightarrow{} 0$) limit, we get average payoff per player as,
\begin{equation}
\underset{\beta \xrightarrow{} 0}{\lim} \langle U \rangle = (2r-p)/2.
\end{equation}
It is due to the reasons we discussed in the previous section wherein all strategy pairs become equiprobable in infinite noise (or zero selection intensity $\beta \xrightarrow{} 0$) limit. In zero noise (high selection intensity $\beta \xrightarrow{} \infty$) limit, since payoffs $r>0$ and $p>0$, $2r$ will always be greater than $-p$. Thus, for zero noise ($\beta \xrightarrow{} \infty$) limit in Eq.~(\ref{eqn:HDavgU}) we get,
\begin{equation}
\underset{\beta \xrightarrow{} \infty}{\lim}\langle U \rangle =2r.
\end{equation}
We get the average payoff per player in zero noise ($\beta \xrightarrow{} \infty$) limit as $2r$, which corresponds to strategy $(provide, provide)$ in the payoff matrix of Eq.~(\ref{eqn:publicgoodsgamematrix}). We get this despite Nash equilibrium being $(free-ride,free-ride)$ for $r<(t/2-p)$. Following what we saw for game magnetization in zero noise ($\beta \xrightarrow{} \infty$) limit, without randomization of strategy, the HD model always gives a payoff corresponding to both players cooperating since the collective payoff of players in thermodynamic limit is largest when all players provide.
\subsection{\label{PG-DE}Results from DE model}
This section will analyze game magnetization and average payoff per player using the DE model.
\subsubsection{\label{PG-DE-Mag}Game magnetisation}
Since payoffs for public goods game obey $a+d=b+c$, we have game magnetisation from Eq.~(\ref{eqn:MagDE}) as,
\begin{equation}
\langle \hat{J_z}^{(1)} \rangle_\beta= \tanh \beta\bigg(\frac{2r-t+2p}{4}\bigg).
\end{equation}
It is the same as that obtained from NE mapping in Eq.~(\ref{eqn:NEmagPublicgoods}). However, the DE model's inconsistencies regarding the average payoff in zero noise ($\beta \xrightarrow{} \infty$) limit become apparent.

\subsubsection{\label{PG-DE-avgU}Average payoff per player}
In order to derive the average payoff per player from the partition function in Eq.~(\ref{eqn:DEpartitionfunction}), we have,
\begin{equation}
Z=e^{\beta 2r}+e^{\beta (r-t/2)}+e^{\beta (r+t/2-p)}+e^{-\beta p}.
\end{equation}
Just as we saw in previous sections, the average payoff per player for the DE model is given by the negative of average thermodynamic energy $\langle E \rangle$,
\begin{align}
\langle U \rangle=&-\langle E \rangle=\frac{\partial \ln Z}{\partial \beta}
\notag \\
=&\frac{\resizebox{.4 \textwidth}{!} {$-pe^{-\beta p}+(r+\frac{t}{2}-p)e^{\beta(r+\frac{t}{2}-p)} +(r-\frac{t}{2})e^{\beta(r-\frac{t}{2})} + (2r)e^{\beta(2r)}$}}{e^{-\beta p}+ + e^{\beta(r+\frac{t}{2}-p)} +e^{\beta(r-\frac{t}{2})} + e^{\beta(2r)}}.
\label{eqn:DEavgU}
\end{align}
For infinite noise, i.e., $\beta \xrightarrow{} 0$ (or, zero selection intensity) limit, we get average payoff per player as,
\begin{equation}
\underset{\beta \xrightarrow{} 0}{\lim}\langle U \rangle = r - p/2.
\end{equation}
Again, this happens as all available strategies become equiprobable for infinite noise limits. On the other hand, for zero noise (i.e., $\beta \xrightarrow{} \infty$ or high selection intensity) limit, the average payoff depends on the relative magnitude of payoffs as
\begin{enumerate}
\item For condition $r > (t/2-p)$, we have $2r > (r+t/2-p), 2r > (r-t/2), 2r >-p $ ( with $ r, t, p > 0$). Thus, we get the average payoff per player in zero noise limit as
\begin{equation}
\underset{\beta \xrightarrow{} \infty}{\lim}\langle U \rangle = 2r.
\end{equation}
We get the average payoff per player corresponding to the payoff of Nash equilibrium strategy $(provide, provide)$ for $r > (t/2-p)$.
\item Next when we consider $r < (t/2-p)$ and for $t > p$, we get average payoff per player in zero noise ($\beta \xrightarrow{} \infty$) limit as
\begin{equation}
\underset{\beta \xrightarrow{} \infty}{\lim}\langle U \rangle=r + \frac{t}{2}-p.
\end{equation}
It does not correspond to the payoff of the Nash equilibrium strategy, which is $(free-ride,free-ride)$ in $r < (t/2-p)$ regime.

\item For $r < (t/2-p)$ and for $t<p$, we get average payoff per player in zero noise ($\beta \xrightarrow{} \infty$) limit as,
\begin{equation}
\underset{\beta \xrightarrow{} \infty}{\lim}\langle U \rangle = r - \frac{t}{2}.
\end{equation}
Again, this does not correspond to the payoff for Nash equilibrium strategy $(free-ride,free-ride)$.
\end{enumerate}
So for $r > (t/2-p)$, average payoff in zero noise ($\beta \xrightarrow{} \infty$) limit corresponds to Nash equilibrium while for $r<(t/2-p)$, it does not. For the DE model, even though game magnetization is in excellent agreement with that from NE mapping, the average payoff per player in zero noise limit is not. It is no surprise since the DE model only concerns maximizing the payoff of a single focal player without regard for the payoff of the neighbouring player. DE model thus gives the average payoff in zero noise limit as the value of the largest payoff in the game matrix.
\subsection{\label{PG-ABM}Results from Agent-based method}
This section will analyze game magnetization and average payoff per player using an Agent-based method. In order to do an Agent-based simulation for the public goods game, we take, as before energy matrix as negative of the payoff matrix of Eq.~(\ref{eqn:publicgoodsgamematrix}) and proceed with the algorithm as described in Sec.~\ref{ABMtheory}.
\begin{equation}
\text{Thus, }E=
\bigg(\begin{array}{c c}
-2r & -r+\frac{t}{2}\\
-r-\frac{t}{2}+p & p
\end{array}\bigg).
\label{eqn:PGagentenergymatrix}
\end{equation}
We also find the average payoff per player in zero noise limit using the algorithm detailed in~\ref{Chicken-ABM-avgU}. The game magnetization obtained via the Agent-based method is shown in Fig.~\ref{fig:three graphs}, and its analysis is done in Sec.~\ref{PG-Analysis-Mag}. The average payoff per player in zero noise limit obtained via the Agent-based method is given in Fig.~\ref{fig:PublicGoodsaverageUcombined}, and its analysis is done in Sec.~\ref{PG-Analysis-AvgU}.

\subsection{\label{PG-Analysis} Analysis of Public goods game}
Herein we analyze game magnetization as well as the average payoff calculated in Sec.~\ref{PG-NE} for NE mapping (NEM), Sec.~\ref{PG-HD} for Hamiltonian dynamics (HD) model, Sec.~\ref{PG-DE} for Darwinian evolution (DE) model and Sec.~\ref{PG-ABM} for agent-based method (ABM) in the context of public goods game.
\subsubsection{\label{PG-Analysis-Mag}Game magnetization}
\begin{figure*}
\centering
\subfloat[]{\includegraphics[height=5.0cm]{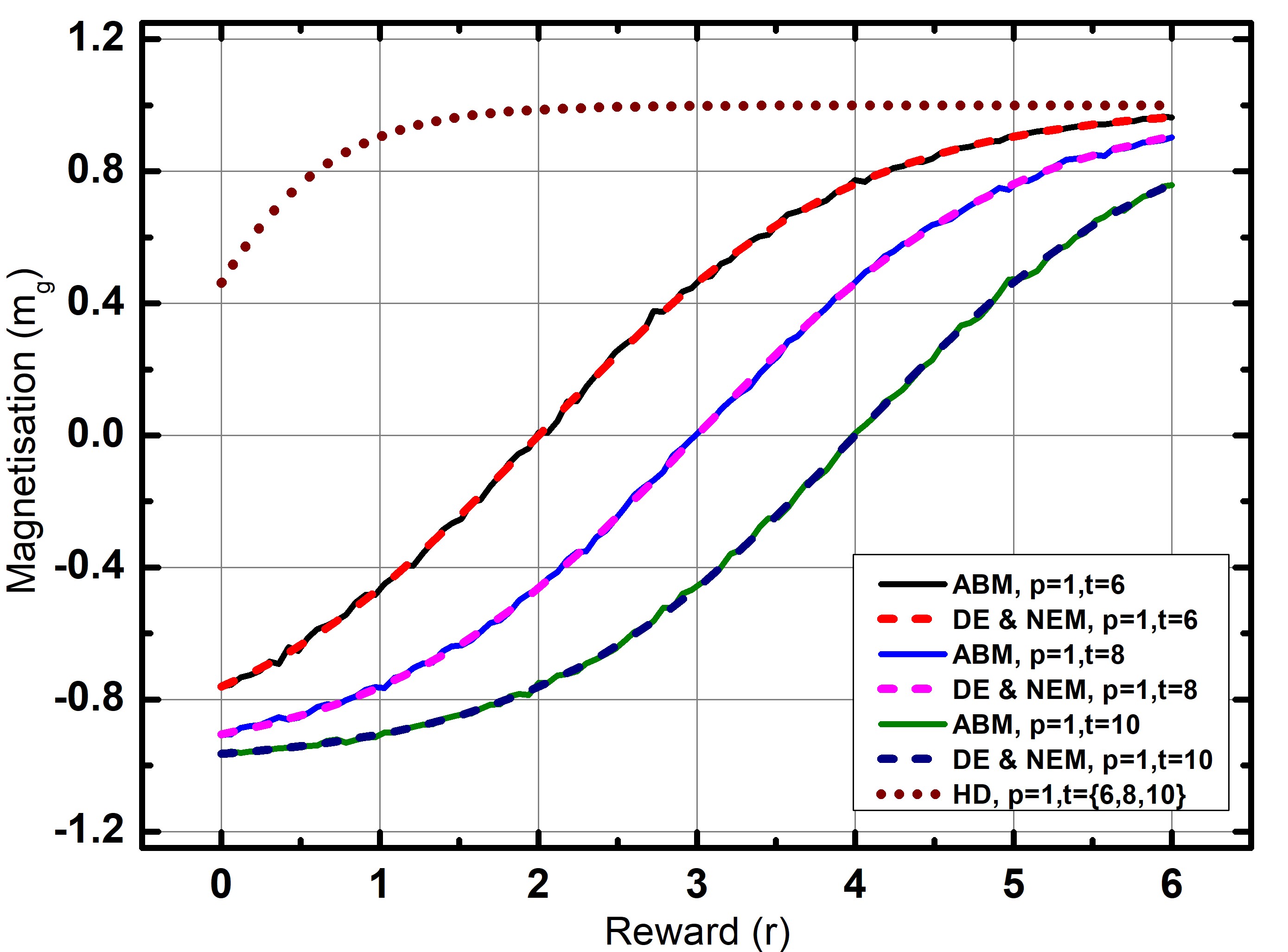}}
\hspace{5mm}
\subfloat[]{\includegraphics[height=5.0cm]{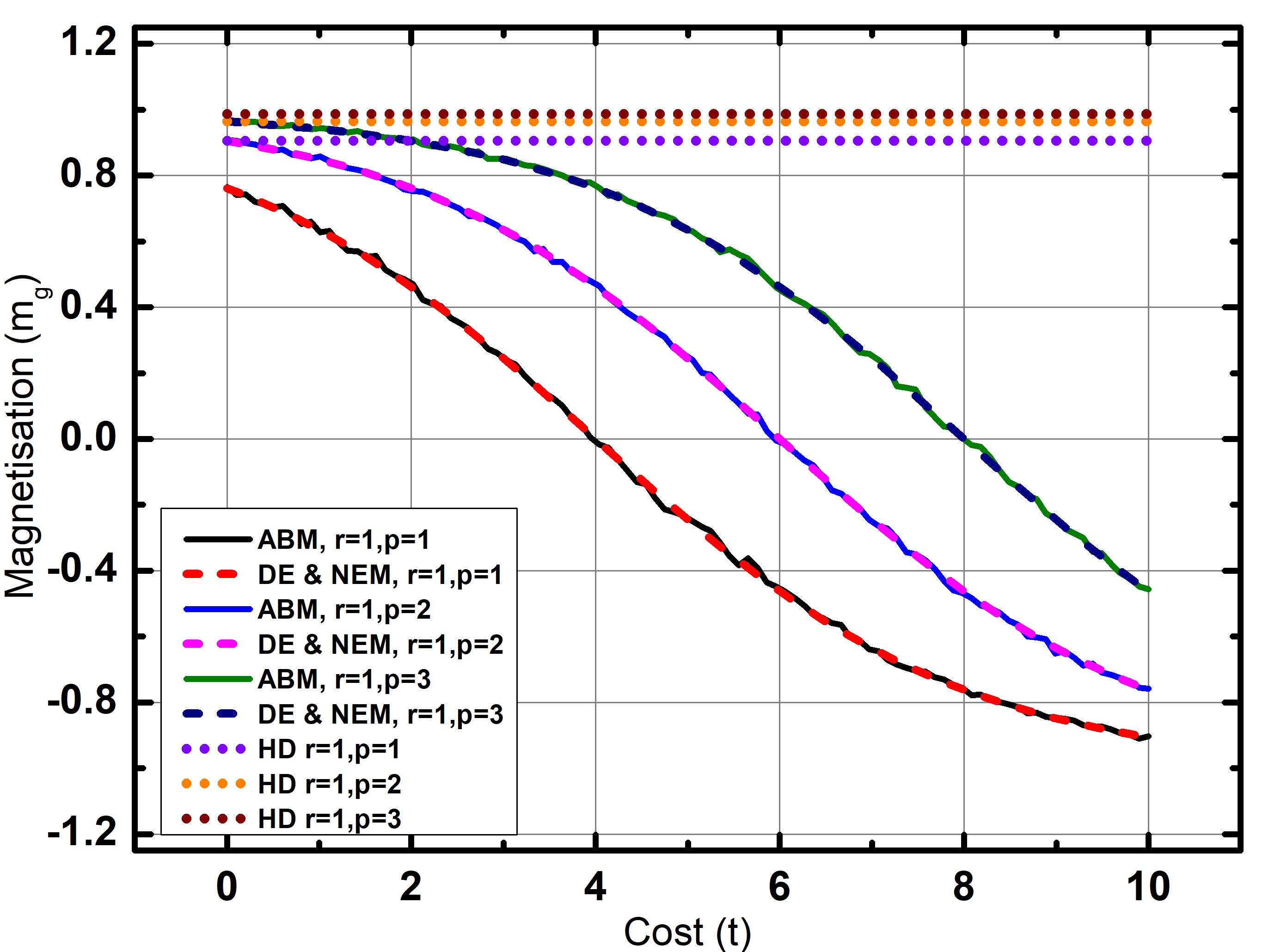}}
\caption{\textbf{(a)} Game magnetization $m_g$ vs. reward $r$ for public goods game as obtained by Hamiltonian dynamics (HD) model, Darwinian evolution (DE) model, NE mapping (NEM) and agent-based method (ABM) for $\beta=1.0$, punishment $p=1.0$ and three different values of cost $t=6.0,8.0,10.0$. \textbf{(b)} Game magnetization $m_g$ vs. cost $t$ for public goods game as obtained by Hamiltonian dynamics (HD) model, Darwinian evolution (DE) model, NE mapping (NEM) and agent-based method (ABM) for $\beta=1.0$, reward $r=1.0$ and three different values punishment $p=1.0,2.0,3.0$.}
\label{fig:three graphs}
\end{figure*}
In this subsection, we will look at game magnetization vs. payoffs for three analytical models and compare them with Agent-based simulation results in the context of public goods games.

In Fig.~\ref{fig:three graphs}(a), we have plotted game magnetization versus reward as obtained by the HD model (dotted line), NE mapping (dashed lines), DE model (dashed lines), and agent-based method (solid line) for $\beta=1.0$ and cost $t=6.0,8.0,1.0$ with punishment fixed at $p=1.0$. For Hamiltonian dynamics, game magnetization is always positive since the collective payoff is largest when all players cooperate. Also, we see that game magnetization is independent of cost $t$. As we have analytically seen, NE mapping and the DE model give the same value for game magnetization. This game magnetization has a transition point at resource value $r=(t/2-p)$ where it goes from the majority being free riders to the majority being providers as $r$ increases. It is also the point where Nash equilibrium for a $2$-player public goods game shifts from $(free-ride,free-ride)$ to $(provide, provide)$. Additionally, game magnetization obtained from the Agent-based method is in excellent agreement with both NE mapping and the DE model.\\
\begin{figure}
\centering
\includegraphics[width=8.5cm]{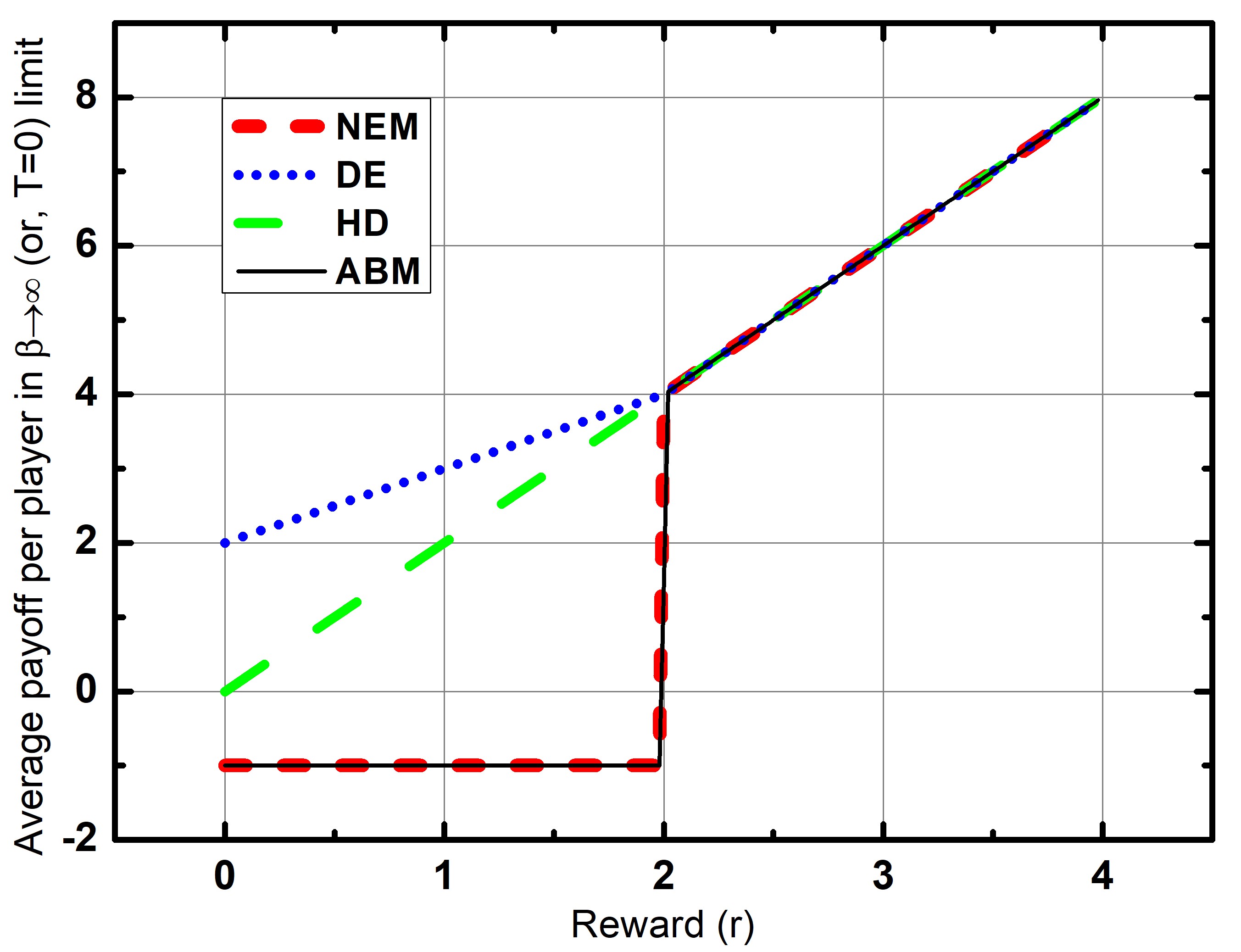}
\caption{$\underset{\beta \xrightarrow{} \infty}{lim} \langle U \rangle$ vs $r$ for public goods game obtained via NE mapping (NEM), Hamiltonian dynamics (HD) model, Darwinian evolution (DE) model and agent-based method (ABM) for punishment $p=1$ and cost $t=6$.}
\label{fig:PublicGoodsaverageUcombined}
\end{figure}
In Fig.~\ref{fig:three graphs}(b), we plot game magnetization against cost $t$ for $\beta=1.0$ with punishment $p=1.0,2.0$ and $3.0$ and value of reward fixed at $r=1.0$. In the case of the HD model, game magnetization is constant throughout the cost range $t$ since it is independent of cost, although it is dependent on punishment $p$. Once again, a positive value for game magnetization is obtained throughout the entire cost range $t$ and disagrees with game magnetization from other analytical models. Agent-based simulation, on the other hand, agrees perfectly with game magnetization obtained from both NE mapping and DE models. For the DE model, however, this is a false positive since a player in the DE model favors the strategy corresponding to the largest payoff in the game matrix without any regard to the payoff of its interacting neighbour. When game payoffs obey the condition $a+d=b+c$- which it does for public goods game- it so happens that the Nash equilibrium strategy (which considers payoffs of both interacting neighbours) always coincides with this "selfish" strategy of the player in the DE model.
\subsubsection{\label{PG-Analysis-AvgU}Average payoff per player}
In Fig.~\ref{fig:PublicGoodsaverageUcombined}, we plot the average payoff per player in zero noise ($\beta \xrightarrow{} \infty$) limit vs reward ($r$) for punishment $p=1$ and cost $t=6$ as obtained via NE mapping (dashed red line), HD model (dashed green line), DE model (dotted blue line) and agent-based method (solid black line). We have a phase transition point with these parameters at $r=(t/2-p)=2$. For NE mapping, we get the average payoff per player in zero noise ($\beta \xrightarrow{} \infty$) limit as $-1$ (i.e., $-p$) for $r < 2$ and $2r$ for $r > 2$- both of these are Nash equilibrium payoffs. Hence NE mapping consistently picks out the Nash equilibrium strategy in zero noise limit and follows that path for the public goods game. For the HD model, the average payoff per player, in zero noise limit, is $2r$ irrespective of punishment $p$ or cost $t$, which corresponds to all players cooperating. It is because the HD model seeks to maximize the combined payoff of all players, and this happens in public goods games when all players cooperate. For the DE model, we get the average payoff per player in zero noise limit as $r+2$ (i.e., $r+t/2-p$) for $r < 2$ and $2r$ for $r > 2$. Thus, the average payoff per player in the DE model predicts Nash equilibrium payoff for $r>2$, but this is only coincidental since the DE model always opts for the largest payoff in the game payoff matrix in $\beta \xrightarrow{} \infty$ limit and $2r$ happens to be that in the range $r>2$. Finally, for the Agent-based method, we see that the average payoff per player in zero noise limit equals Nash equilibrium payoff, i.e., $-1$ for $r < 2$ and $2r$ for $r > 2$. Hence average payoff per player for NE mapping and Agent-based method match exactly in zero noise limit.

\section{Conclusion}
\begin{table}[h]
\centering
\begin{tabular}{|c|c|c|c|c|}
\hline
\textbf{} & $\beta$ limits & NE mapping & DE & ABM \\ \hline
\multirow{2}{*}{$m_g$} & $\beta \xrightarrow{} 0$ & $0$ & $0$ & $0$ \\ \cline{2-5}
& $\beta \xrightarrow{} \infty$ & $0$ & \textcolor{red}{$\mathbf{1}$} & $0$ \\ \hline
\multirow{2}{*}{$\langle U \rangle$} & $\beta \xrightarrow{} 0$ & $-s/4$ & $-s/4$ & $-s/4$ \\ \cline{2-5}
& $\beta \xrightarrow{} \infty$ & $0$ & \textcolor{red}{$\mathbf{r}$} & $0$ \\ \hline
\end{tabular}
\caption{$m_g$ and $\langle U \rangle$ in zero noise ($\beta \xrightarrow{} \infty$) and infinite noise ($\beta\xrightarrow{} 0$) limits for the Hawk-Dove game as obtained by NE mapping, DE model and agent-based method (ABM). The incorrect results obtained are highlighted in red.}
\label{Chickentable}
\end{table}

\begin{table*}[t]
\scriptsize
\centering
\begin{tabular}{|c|c|c|c|c|c|}
\hline
\textbf{} & $\beta$ limits & NE mapping & DE & HD & ABM \\ \hline
\multirow{2}{*}{$m_g$} & $\beta \xrightarrow{} 0$ & 0 & 0 & 0 & 0 \\ \cline{2-6}
& $\beta \xrightarrow{} \infty$ & $\begin{cases} +1 & , r > t/2 - p\\
-1 & , r < t/2-p \end{cases}$ & $\begin{cases} +1 & , r > t/2 - p\\
-1 & , r < t/2-p \end{cases}$ & \textcolor{red} {$\mathbf{+1}$} & $\begin{cases} +1 & , r > t/2 - p\\
-1 & , r < t/2-p \end{cases}$ \\ \hline
\multirow{2}{*}{$\langle U \rangle$} & $\beta \xrightarrow{} 0$ & $r - p/2$ & $r - p/2$ & $r - p/2$ & $r-p/2$ \\ \cline{2-6}
& $\beta \xrightarrow{} \infty$ & $\begin{cases} 2r & , r > t/2 - p\\
-p & , r < t/2-p \end{cases}$ & $\begin{cases} 2r & ,r > t/2 - p\\
\textcolor{red} {\textbf{r + t/2 - p}} &, r < t/2 - p, t > p\\
\textcolor{red} {\textbf{r - t/2}} & , r < t/2 - p, t < p\end{cases}$ & \textcolor{red} {\textbf{2r}} & $\begin{cases} 2r & ,r > t/2-p\\
-p & , r <t/2 - p \end{cases}$ \\ \hline
\end{tabular}
\caption{Game magnetization and average payoff per player in the zero noise ($\beta \xrightarrow{} \infty$) and infinite noise ($\beta \xrightarrow{} 0$) limits for public goods game as obtained by NE mapping, DE model, HD model and Agent based method (ABM). The incorrect results obtained are highlighted in red.}
\label{Publicgoodstable}
\end{table*}

In the Hawk-Dove game, NE mapping accounted for the mixed Nash equilibrium of the Hawk-Dove game as indicated by the phase transition point of the game magnetization curve going from negative to positive at $r=s/2$ at finite $\beta$ values. In zero noise limit, game magnetization and average payoff per player indicate that players adopt either of the two pure Nash equilibria strategy pairs- $(Hawk, Dove)$ or $(Dove, Hawk)$. The game magnetization obtained via NE mapping is very close to that obtained via Agent based method. Game magnetization, as well as the average payoff in zero noise limit obtained via DE model, on the other hand, {diverges} significantly from both NE mapping and an agent-based method {on top of being} completely uncharacteristic of the Hawk-Dove game. Table~\ref{Chickentable} lists the infinite noise ($\beta \xrightarrow{} 0$) and zero noise ($\beta \xrightarrow{} \infty$) limits for $m_g$ and $\langle U \rangle$ in case of NE mapping, DE model and agent-based method. Table~\ref{Chickentable} clearly shows excellent agreement between NE mapping and agent-based method in these limits. Additionally, Table~\ref{Chickentable} shows the inconsistency of DE model \textit{vis-a-vis} both NE mapping and agent-based method. DE model fails because it seeks to maximize the payoff of only a single player with no regard to the strategy of neighbouring players.\\

In the Public goods game, magnetization is obtained from NE mapping and DE model match (since the game satisfies condition $a+d=b+c$). We also see transition points where one expects them to when looking at Nash equilibrium of $2$ player public goods game. The strategy adopted by most players in a game always matches its Nash equilibrium strategy. Additionally, game magnetization obtained via Agent based method agrees with this exactly. However, game magnetization obtained via the HD model disagrees with that obtained from the other models. Game magnetization of the HD model indicates the majority of players always provide with no regard to Nash equilibrium strategy, which is $(free-ride,free-ride)$ for $r < (t/2-p)$.
Regarding average payoff per player in zero noise limit, NE mapping and Agent based method gives results that agree perfectly with the Nash equilibrium strategy. DE model, on the other hand, disagrees with the average payoff in zero noise limit, with both NE mapping and Agent based method and Nash equilibrium strategy payoff. Finally, the HD model always gives an average payoff in zero noise ($\beta \xrightarrow{} \infty$ or high selection intensity) limit as corresponding to the $(provide, provide)$ strategy of the game. It is because HD seeks to maximize the collective payoff of all players, which happens when all players in social dilemma settings choose to provide. Table~\ref{Publicgoodstable} lists the infinite noise ($\beta \xrightarrow{} 0$) and zero noise ($\beta \xrightarrow{} \infty$) limits of $m_g$ and $\langle U \rangle$ in case of NE mapping, HD model, DE model, and an agent-based method. Table~\ref{Publicgoodstable} clearly shows the agreement between NE mapping and agent-based method in these limits. Table~\ref{Publicgoodstable} also shows the inconsistency of the HD model and DE model \textit{vis-a-vis} NE mapping and ABM. Just as the DE model failed for the Hawk-Dove game, it also fails here. The DE model maximizes a single player's payoff while ignoring the interact neighbours' payoff considerations. Our study suggests the NE mapping approach as the only accurate analytical model available to study $2-$ player and $2-$strategy games in the thermodynamic limit. Both DE and HD models are inaccurate and should not be used to analyze the thermodynamic limit of similar games.

\section*{Acknowledgments}
The grants have funded this study: (1) Josephson junctions with strained Dirac materials and their application in quantum information
processing, SERB Grant No. CRG/20l9/006258, and (2) Nash equilibrium versus Pareto optimality in N-Player games, SERB MATRICS Grant No.
MTR/2018/000070.
\section*{Author declarations}
\subsection{Conflict of Interest} The authors have no conflicts to disclose.
\subsection{Data availability statement} The data supporting this study's findings are available within the article and Appendix.
\subsection{Author contributions} CB gave the initial idea for the project, edited the paper, and wrote the code in Python 3. He also administered the project. He provided the key references and instructed the other author to initiate him into the project. CB replied to the reviewer's queries and made changes to the manuscript as required. AKUM wrote the first draft of the paper, made the graphs and did the calculations, and wrote code in Python 2.7.

\onecolumngrid
\begin{appendix}
\section{Agent based simulation}
The python3 code we used for finding the magnetization vs. reward graph for the Hawk-Dove game (see FIG. 1) using Agent-based simulation is given below.
\lstset{language=Python}
\lstset{frame=lines}
\lstset{label={lst:code_direct}}
\lstset{basicstyle=\footnotesize}
\begin{lstlisting}
import numpy as np
import matplotlib.pyplot as plt
a=np.random.rand(1000)
#1-D string of 10,000 players
PU=np.linspace (0.0,4.0,100)
#Domain of punishment where we plot magnetization
for i in range(0,len(a)):
if(a[i]<0.5):
a[i]=int('0')
else:
a[i]=int('1')
m=[]
s=4.0
T=1.0 #
for r in PU:
E=[[s,-r],[ r,0 ]] #Energy matrix
for k in range(0,1000000):
#10 million iterations; average 1000 iterations per player
i=np.random.randint(len(a))
#Randomly choosing a player
p=np.random.rand()
#Random value of p between 0 and 1
if((p)<=(1/(1+np.exp(-(E[int(a[i])][int(a[(i+1)\%len(a)])]
-E[int(a[i]+1)\%2][int(a[(i+1)\%len(a)])])/T)))):
a[i] = (int(a[i]+1)\%2)
#Flipping the strategy when $p < 1/(1+exp(\beta\Delta E))$
m=np.append(m,(len(a)-2*sum(a))*1.0/len(a))
#Magnetisation values
plt.plot(PU,m,label='Agent based model')
\end{lstlisting}
The code is the same for simulating the public goods game, except for the energy matrix, where $E$ is negative of the public goods game payoff matrix.
\end{appendix}
\end{document}